\documentclass[a4paper,10pt,hyper]{JHEP3}
\usepackage{amsfonts,latexsym,graphicx,epsfig,amssymb,amsmath,mathrsfs,multirow}
\usepackage{color}

\newcommand{\ma}[1]{\mbox{$\mathcal{#1}$}}

\newcommand{\tr}{{\rm Tr}}
\newcommand{\ti}{\tilde}

\newcommand{\SL}{{\rm SL}$(8, \mathbb R)$}
\newcommand{\SP}{{\rm Sp}$(56, \mathbb R)$}

\title{
Classification and stability of vacua in maximal gauged supergravity
}
\author{
Hideo Kodama\\
Theory Center, KEK, Tsukuba 305-0801, Japan;\\
Department of Particles and Nuclear Physics,
The Graduate University for Advanced Studies, Tsukuba 305-0801, Japan\\
\email{Hideo.Kodama@kek.jp}
}
\author{
Masato Nozawa\\
Theory Center, KEK, Tsukuba 305-0801, Japan\\
\email{nozawam@post.kek.jp}
}

\abstract{
This article presents a systematic study of critical points
for the ${\rm SL}(8, \mathbb R)$-type gauging in
four dimensional maximal gauged supergravity.
We determine all the possible vacua for which
the origin of the moduli space becomes a critical point via
${\rm SL}(8, \mathbb R)$ transformations.
We formulate a new tool which enables us to find analytically the mass spectrum of
the corresponding vacua in terms of eigenvalues of the embedding
tensor.
When the cosmological constant is nonvanishing,
it turns out that many vacua obtained by the dyonic embedding
admit a single deformation parameter of the theory,
in agreement with the results of the
recent paper by Dall'Agata, Inverso and Trigiante~\cite{Dall'Agata:2012bb}.
Nevertheless, it is shown that the resulting mass spectrum is
independent of the deformation parameter and can be classified
according to the unbroken gauge symmetry at the vacua, rather than the
underlying gauging.
We also show that the generic Minkowski vacua exhibit instability.
}

\preprint{KEK-TH-1582, KEK-Cosmo-105}

\begin{document}


\section{Introduction}

The maximal supergravity has played a distinguished role
in the development of string/M-theory.
Although the maximal supergravity fails to describe our realistic chiral world,
lots of attentions have been paid to this theory mainly due to the hope of ultraviolet
finiteness. Thanks to the high degree of supersymmetry,
the particle spectrum is unified into a single supermultiplet and
there is no freedom to couple additional matter fields.
The only known deformation of maximal supergravity is to gauge the theory
by promoting the abelian gauge fields to the nonabelian ones.
The gauging procedure gives rise to the scalar potential,
as well as the fermionic mass terms.
Recently the gauged supergravity theories have been intensively
studied in the context of flux compactifications, the gauge/gravity duality
and also the condensed matter physics applications.

The original $N=8$ ungauged supergravity was  constructed by
Cremmer and Julia via a toroidal compactification of
eleven dimensional supergravity~\cite{CJ}.
de Wit and Nicolai provided the first example of
maximal gauged supergravity by gauging 28 vector fields to have
an ${\rm SO}(8)$ gauge invariance, based on the formalism of
$T$-tensor~\cite{DN}. This theory has a simple higher dimensional
origin, since it is obtained  by a dimensional reduction of eleven dimensional
supergravity on a seven sphere~\cite{DN2}.
Later on, some noncompact gaugings were found to be possible without
giving rise to ghost~\cite{Hull}.  Subsequently, these types of gaugings have provided
a variety of nontrivial vacua.  It is then important to explore
which types of gaugings are consistently realizable. However, this is not an easy
task since viability is sensitive to the choice of (possibly
non-semisimple)  gauge groups and their embeddings~\cite{Cordaro:1998tx,Hull:2002cv}.
Even if the consistent gauging is assigned, the extremalization of scalar potential is a formidable
task in a general setting, since 70 scalar fields appear in the theory.

Thus, the vacuum hunting so far has been mainly focused upon the truncated sectors
where only a few invariant scalars survive.
This strategy has been active during the past 25 years~\cite{Hull,Warner:1983vz,Hull:1984ea,Ahn:2001by,Ahn:2002qga,Kallosh:2001gr,Bobev:2010ib,Bobev:2011rv,Fischbacher:2010ec}.
In this traditional approach we need to choose the particular gauging, compute the
scalar potential and then scan the moduli space of critical points
of the potential. Specifically,
de Sitter (dS) extrema have been found for ${\rm SO}(4, 4)$
and ${\rm SO}(5,3)$ gaugings by confining to the ${\rm SO}(p)\times {\rm SO}(q)$
invariant scalar. At these vacua spontaneous supersymmetry
breakings occur, hence they may be relevant for the early stages of the
universe. Although this approach offers a concise way to find vacua,
it does not reveal more than the invariant scalars of
specific subgroup.
For example, despite the fact that  all singlet scalars of
 ${\rm SU}(4)^-\subset {\rm SO}(8)$ invariant sector are stable,
non-singlet scalars do indeed have instabilities~\cite{Bobev:2010ib}.
It is therefore desirable to address the systematic
survey of viable gaugings, scanning vacua and full stability thereof.

We have recently witnessed two progresses in this line of research.
One is the development of a new computational tool to find vacua~\cite{Fischbacher:2009cj}.
A dozen of new critical points have been discovered numerically.
Remarkably it was worked out that some nonsupersymmetric vacua
are perturbatively stable. These intensive works
indicate the possibility of an abundant variety of vacua with
potential phenomenological applications.
For instance, the anti-de Sitter (AdS) vacua are expected to be dual to the
nontrivial conformal fixed points in the dual field theory.

Another development is the formulation based on the embedding
tensor~\cite{Nicolai:2000sc,deWit:2002vt,de Wit:2007mt}.
The embedding tensor specifies
how to embed the gauge group into the duality group.
Using this formalism, all different gaugings can be described in a
covariant manner and admissible gaugings can be
characterized group-theoretically.

Under these  circumstances, Dall'Agata and Inverso
utilized the homogeneity of the scalar coset space to determine the
complete mass spectrum of 70 scalar fields for some gaugings~\cite{DI}
(see~\cite{Dibitetto:2011gm} for an early study in half maximal supergravity).
Instead of viewing the scalar potential as nonlinear functions of
seventy scalar fields, it may be identified as
a quadratic function of the embedding tensor. Since the
scalar coset $E_{7(7)}/{\rm SU}(8)$ is homogeneous, any point
can be brought to the origin by the $E_{7(7)}$ isometry,
which acts also on the embedding tensor. Hence the critical point of the
scalar potential can be mapped to the origin, at the price of
varying the embedding tensor. Namely we can explore the
possible gaugings, critical points and their mass spectrum
at the origin of the scalar manifold, where the governing equations can
be analyzed algebraically.
See e.g.,~\cite{Borghese:2011en,Dall'Agata:2012cp,Borghese:2012qm} for some related works
by this method.

The aim of this paper is to deepen our understanding of
vacuum structure in maximal supergravity.
We make a systematic study of vacua which can be moved to the origin of
moduli space via \SL~transformations and
address some issues unresolved in~\cite{DI}.
We give a new tool which enables us to trace analytically the vacuum stability
without resorting numerics or annoying diagonalization of
$70\times 70$ mass matrices.
We conclude that apart from the Minkowski vacua, the mass spectrum is determined by the residual
gauge symmetry, rather than the gauging itself.
In the meanwhile, the Minkowski vacua are shown to admit intricate mass spectra
and possess instabilities in general.

This paper is organized as follows.
In the next section we succinctly  describe the embedding tensor formalism
and fix our notations. Sections~\ref{sec:electric}
and~\ref{sec:dyonic} are devoted to the discussion
of vacuum classifications and mass spectra.
Finally, we conclude with some future prospects in section~\ref{sec:conclusion}.

\section{Maximal gauged supergravity}
\label{sec:maxSUGRA}

In this section we will briefly discuss the gauging
of maximal supergravity and fix our notations.
In the  maximal supergravity, the scalar manifold is described by
the $E_{7(7)}/{\rm SU}(8)$ nonlinear sigma model. The $E_{7(7)}$
acts on the coset representative as isometries, while
acts on the gauge fields as global symmetries.
We choose a subgroup of $E_{7(7)}$ and promote it to a local symmetry.
In order to keep the supersymmetry,
this deformation gives rise to a nontrivial scalar potential, by which
70 scalar fields may get stabilized by acquiring mass. For this purpose,
the embedding tensor formalism is of help, since it allows one to trace all equations
formally in an $E_{7(7)}$ covariant fashion.
Refer to the original paper~\cite{de Wit:2007mt} for a more rigorous discussion.

\subsection{Embedding tensor formalism}

Since the gauge group is a subgroup of $E_{7(7)}$,
its generators $X_M$ can be expressed in terms of the
generators, $t_\alpha $,  of $E_{7(7)}$ as
\begin{align}
X_M ={\Theta _M}^\alpha t_\alpha \,,
\label{N8_embeddingtensor}
\end{align}
where $\alpha =1,..., 133$ and
$M=1,...,56$. The gaugings are encoded into the
real embedding tensor  $\Theta_M{}^\alpha $
belonging to the ${\bf 56}\times{\bf 133}$
representation of $E_{7(7)}$. It
specifies which generators of the duality group
to be chosen as generators of gauge group.
A major advantage of adopting the embedding tensor is that it allows us
to keep the entire formulation in a duality covariant way.
In terms of $X_M$ the symmetry can be made local by introducing
gauge covariant derivative,
\begin{align}
 \partial_\mu ~~\to ~~ D_\mu =\partial_\mu -g A_\mu{} ^M X_M \,,
\end{align}
where $g$ is the coupling constant.

The embedding tensor must satisfy linear and quadratic constraints
in order to ensure the consistent gaugings.
The quadratic constraint equation requires that the
embedding tensor should be invariant under the gauge group,
\begin{align}
 {C_{MN}}^\alpha :=f_{\beta\gamma }{}^\alpha \Theta_M{}^\beta
 \Theta _{N}{}^\gamma +(t_\beta )_N{}^P \Theta_M{}^\beta \Theta_P
 {}^\alpha=0 \,,
\label{N8_ET_quadratic}
\end{align}
where $f_{\alpha \beta }{}^\gamma $ denotes the structure constants of
$E_{7(7)}$, i.e.,
$[t_\alpha ,t_\beta ]={f_{\alpha\beta }}^\gamma t_\gamma $.
Equation~(\ref{N8_ET_quadratic}) implies the closure condition
$[X_M, X_N]=-X_{MN}{}^PX_P$ of the gauge algebra.
It should be stressed that (\ref{N8_ET_quadratic}) involves a nontrivial
information upon the symmetrization in $(M,N)$.

In addition to the quadratic relation,
the supersymmetry imposes the linear constraint upon the embedding
tensor. The concrete form of this constraint depends on the spacetime
dimensionality and supercharges. In the present case,
the embedding tensor is subject to the following restriction~\cite{deWit:2002vt,de Wit:2007mt},
\begin{align}
 (t_\alpha )_M{}^N \Theta_N{}^\alpha =0 \,, \qquad
(t_\beta t^\alpha  )_M{}^N \Theta_N{}^\beta =-\frac 12 \Theta_M{}^\alpha
 \,.
\label{N8_CP_linearconst}
\end{align}
Here and in what follows, the indices $\alpha , \beta,... $ are raised and lowered by the
$E_{7(7)}$ Cartan-Killing metric $\eta_{\alpha\beta }=\tr (t_\alpha t_\beta) $.
Although the embedding tensor {\it a priori} takes values in the tensor product
${\bf 56}\times {\bf 133}={\bf 56}+{\bf 912}+{\bf 6480}$,
the above constraint amounts to requiring the embedding tensor
to belonging to the {\bf 912} representation~\cite{deWit:2002vt,de
Wit:2007mt}.\footnote{
If one gauges the trombone symmetry, the {\bf 56} representations can be
excited~\cite{LeDiffon:2011wt}. }
From equation~(\ref{N8_CP_linearconst}) one may derive
$X_{(MNP)}=X_{M[NP]}={X_{MN}}^M=0$, namely, the gauge group must be unimodular.
Each solution to the above set of constrains gives rise to
a viable gauging.

In the present paper, we are interested in a gauge group
embedded into the standard \SL~subgroup of $E_{7(7)}$.
The branching rules into \SL~representations are
\begin{align}
{\bf 56}\to {\bf 28}+{\bf 28}' \,, \qquad
{\bf 133}\to {\bf 63}+{\bf 70}\,, \qquad
 {\bf 912}\to {\bf 36}+{\bf 420}+{\bf 36}'+{\bf 420}'\,.
\end{align}
Since the
embedding tensor lives in ${\bf 912}$ representation,
the relevant branchings are given by
\begin{align}
\begin{array}{c|cc}
  &{\bf 28}  & {\bf 28}' \\
\hline
{\bf 63} & ~{\bf 36}+{\bf 420}
& ~{\bf 36}'+{\bf 420}'
\\
{\bf 70} & {\bf 420}' & {\bf 420}
 \end{array}
\end{align}
The representations {\bf 420} and ${\bf 420}'$ appear not only in
 representations arising from the adjoint representation
{\bf 63} of \SL, but also in those coming from {\bf 70}.
Hence, for the embeddings in \SL, the embedding tensor has to belong to
{\bf 36} and/or ${\bf 36}'$,  on which we
will concentrate in the rest of the paper.

The scalar potential arises from the $O(g^2)$ corrections of
supersymmetry transformations.
In terms of $X_{M}$, it is given by~\cite{de Wit:2007mt}
\begin{align}
V =\frac{g^2}{672}\left({X_{MN}}^R{X_{PQ}}^SM^{MP}M^{NQ}M_{RS}+7{X_{MN}}^Q{X_{PQ}}^N
 M^{MP}\right) \,,
\label{N8_potential}
\end{align}
where $M_{MN}$ is a real and symmetric matrix
with the inverse $M^{MN}$ and
defined by
\begin{align}
 M= L\cdot{}^T\!  L \,, \qquad M_{MN}=(M)_{MN}\,.
\end{align}
Here $L=L(\phi)$ is the coset representative in the \SP~representation.
From the higher dimensional point of view,
the four dimensional scalar potential encodes the internal geometry
and the flux contributions.
For generic gaugings, the potential is unbounded both below and above,
and fails to have any extrema.

For later convenience, let us recapitulate some coset representations.
Cremmer and Julia introduced the ${\rm Usp}(56)$ representation,
in which the diagonal element of $E_{7(7)}$ algebra
is ${\rm SU}(8)$~\cite{CJ}.
In the ${\rm Usp}(56)$ representation
the coset representatives
take the form,
\begin{align}
L(\phi) _{\underline M}{}^{\underline N}=\exp
\left(
\begin{array}{cc}
0 & \phi_{ijkl} \\
\phi^{ijkl} & 0
\end{array}
\right) \,, \qquad \phi_{ijkl}=\phi_{[ijkl]}=\eta (\star \bar\phi)_{ijkl}\,,
\end{align}
where the underlined indices refer to
${\bf 28}+\overline{\bf 28}$ of ${\rm SU}(8)$, and
$\eta=\pm 1$ corresponds to the chirality of
the spinor representation of ${\rm SO}(8)$ below.
Here $i,j,...$ are $\mathbf 8$ and $\bar{\bf 8}$
of ${\rm SU}(8)$, and are raised and lowered
via complex conjugation, as usual.
The change of basis can be done
via gamma matrices in the real Weyl spinor representation of ${\rm SO}(8)$,
\begin{align}
L_{\underline M}{}^{\underline N}
=\ma S_{\underline M}{}^{P}L_P{}^Q (\ma S^{-1})_Q{}^{\underline N}\,,
\qquad
\ma S_{\underline M}{}^N =\frac{\rm i}{4\sqrt 2}
\left(
\begin{array}{cc}
\Gamma_{ij}{}^{ab} & {\rm i}\Gamma_{ijab}  \\
\Gamma^{ijab} & -{\rm i}\Gamma^{ij}{}_{ab}
\end{array}
\right)\,,
\label{N8_CP_SpUSp_indicestr}
\end{align}
where
$(\Gamma_{ij})^{ab}=(\Gamma^{ab})_{ij}=:\Gamma^{ab}_{ij}$,
and there is no need to distinguish their upper and lower indices.
In particular,
we denote by $\ma V$ to describe the coset representative in a mixed
basis,
\begin{align}
 \ma V_M{}^{\underline N} = L_M{} ^P(\ma S^{-1})_P{}^{\underline N} \,.
\label{N8_CP_mixV}
\end{align}

\subsection{Mass matrix}

The seventy scalars parametrize the  homogeneous
(and moreover symmetric) coset space
$E_{7(7)}/{\rm SU}(8)$. The homogeneity means that
every point on the (Riemannian) manifold can be mapped into any other point via
a global transformation (isometry). In other words, the manifold admits the transitive
group of motions.

What is important here is that the scalar potential is invariant
under the simultaneous transformations of the coset representative and
of the embedding tensor.
Indeed, the potential depends on a single tensorial combination $L^{-1}\Theta $.
To see this, let us define
\begin{align}
 \ti \Theta _M{}^\alpha t_\alpha:=(L^{-1})_M{}^N{\Theta _N}^\alpha
 L^{-1}t_\alpha L \,.
\end{align}
This is the analogue of $T$-tensor in the
\SP~representation. In terms of $\ti \Theta_M{}^\alpha $,
the potential~(\ref{N8_potential}) can be expressed as
\begin{align}
V
= \frac{g^2}{672}\ti \Theta_M{}^\alpha \ti \Theta_M{}^\beta
(\delta_{\alpha\beta }+7 \eta_{\alpha \beta }) \,,
\label{N8_CP_potti}
\end{align}
where
\begin{align}
{\rm Tr}(t_\alpha t_\beta ^\dagger )=\delta_{\alpha\beta } \,, \qquad
{\rm Tr}(t_\alpha t_\beta )=\eta_{\alpha \beta } \,.
\end{align}
In this form, one notices that
the potential depends (quadratically) only on
$\ti \Theta_M{}^\alpha $, as we desired to show. Since any point on the scalar manifold can
be mapped to any other point, the optimal setup is to move the
critical point to the origin, where
$L(O) =\mathbb I_{56}$. At the origin, the extremum condition
amounts to the quadratic conditions on $\Theta_M{}^\alpha $.

To take the first variation of the potential, we first note that the
coset representative can be written as
\begin{align}
L=L(O)\exp \left(-\phi^\rho t_\rho \right)
=L(O)\left[\mathbb I_{56} -\phi^\rho t_\rho
 +\frac{1}{2}\phi^\rho\phi^\sigma  t_\rho t_\sigma +\cdots \right]\,,
\end{align}
where indices $\rho $ and $\sigma $ refer exclusively to 70 noncompact elements of
$\mathfrak e_{7(7)}$ and $\phi^\rho$ denotes the physical scalars.
This implies that
\begin{align}
 \partial_\rho L=-L t_\rho \,, \qquad
\partial_\sigma \partial_\rho L =L t_{(\rho }t_{\sigma )} \,,
\label{N8_CP_dL}
\end{align}
hold at the origin ($\phi^\rho =0$).
Using analogous relations for the derivatives of $L^{-1}$,
the first derivative of $V$ is obtained as
\begin{align}
 \partial_\rho V=\frac{g^2}{336} \left[t_{\rho M}{}^N\ti
 \Theta_M{}^\alpha  \Theta _N{}^\beta(\delta_{\alpha\beta }+7
 \eta_{\alpha \beta })+ \Theta_M{}^\alpha  \Theta_M{}^\beta
 {f_{\rho\beta }}^\gamma \delta _{\alpha \gamma }
\right]\,.\label{N8_CP_dV}
\end{align}
At the origin, $\partial_\rho V=0$ imposes a quadratic
restriction upon $\Theta_M{}^\alpha $, which should be
combined to be solved with  (\ref{N8_ET_quadratic}) and
(\ref{N8_CP_linearconst}). It turns out that we can
scan the critical points and underlying gaugings at the same time,
as demonstrated in~\cite{DI}.

We can furthermore  discuss the mass spectrum at the same time.
The second derivatives of the potential at the origin can be similarly computed to
give
\begin{align}
\partial_\sigma \partial_\rho V =\frac{g^2}{336}&
\left[(t_{(\rho}t_{\sigma )} +{}^T\! t_{(\rho }t_{\sigma )} )_M{}^N \Theta_M{}^\alpha
 \Theta_N {}^\beta (\delta_{\alpha \beta }+7 \eta_{\alpha \beta })
\right.\nonumber \\
& \left. ~~
+ \Theta_M{}^\alpha \Theta_M{}^\beta (-f_{\alpha( \rho }{}^\gamma
 f_{\sigma) \beta }{}^\delta \delta_{\gamma \delta } +(f_{(\rho} f_{\sigma)} )_\alpha {}^\gamma
 \delta_{\beta\gamma })\right. \nonumber\\
& \left. ~~ +2 (t_{\rho M}{}^N {f_{\sigma \alpha }}^\gamma +{t_{\sigma
 M}}^N{f_{\rho \alpha }}^\gamma )\delta_{\beta\gamma }
 \Theta_M{}^{(\alpha } \Theta_N{}^{\beta )}
\right]\,. \label{N8_CP_ddV}
\end{align}
In order to reduce (\ref{N8_CP_ddV}) to a more tractable form,
we rely on the observation
$[{}^T\! t_\alpha , t_\beta ]\in \mathfrak{e}_{7(7)}$,
which suggests that
there exist constants ${c_{\alpha\beta}}^\gamma $
such that
$[{}^T\!t_\alpha , t_\beta ]={c_{\alpha\beta }}^\gamma t_\gamma$.
Applying the Jacobi identity to (${}^T\!t_\alpha, t_\beta ,t_\gamma $),
we have
\begin{align}
c_{(\rho \sigma )}{}^\gamma f_{\gamma \alpha }{}^\delta \delta _{\beta
 \delta }=-c_{\beta (\rho }{}^\delta c_{\sigma )\alpha }{}^\gamma
 \eta_{\delta \gamma }
 -f_{\alpha (\rho  }{}^\gamma {f_{\sigma) \beta }}^\delta \delta
 _{\gamma \delta } \,.
\end{align}
Using this relation,
a simple computation shows that
(\ref{N8_CP_ddV}) can be cast into
\footnote{We can find a basis of $E_{7(7)}$ in which
$c_{(\alpha\beta)}{}^\gamma$ vanishes when both $\alpha$ and $\beta$
correspond to compact directions or to non-compact directions. 
However, in the other mixed cases, $c_{(\alpha\beta)}{}^\gamma$ does not vanish in general.}
\begin{align}
\left. \partial_\rho \partial_\sigma V\right|_{\phi=0}= (M^2 )_{\rho \sigma }
+\frac12 c_{(\rho\sigma)}{}^\gamma\partial_\gamma V \,,
\label{ddVMsq}
\end{align}
where
$M^2 $ describes the mass matrix at the extrema,
\begin{align}
 (M^2)_{\rho \sigma } :=\frac{g^2 }{168} &\left[(s_{(\rho }s_{\sigma
 )})_M{}^N {\rm Tr}(X_M\,{}^T\!X_N+7X_MX_N)+2 (s_{(\rho })_M{}^N{\rm
 Tr}(s_{\sigma )}[X_{(M}, {}^T\!X_{N)}]) \right.
\nonumber \\ &\left. \quad -{\rm Tr}([s_{(\rho }, X_M][s_{\sigma
 )}, {}^T\! X_M])\right]\,,
\label{N8_CP_massmatrix}
\end{align}
with
\begin{align}
 s_\rho  :=\frac 12 (t_\rho  +{}^T \!t_\rho )\,.
\end{align}
Since (\ref{N8_CP_massmatrix}) is evaluated at the origin,
we can also view $s_\rho$ as dynamical variables rather than
constant $56\times 56 $ symmetric matrices.  In the following
discussion, we will treat $s_\rho $ as linear fluctuations of 70 scalars
around the origin of moduli space.

\section{Electric gaugings}
\label{sec:electric}

We first discuss the case in which the gauge group is
contained in the \SL~electric frame.
This is the simplest setup where the relation to the
flux compactification is clear~\cite{D'Auria:2005rv,Dall'Agata:2007sr,Dibitetto:2012ia}.
This type of gaugings corresponds to the
${\rm CSO}(p,q,r)$ gaugings.

Since the embedding tensor is described by the
${\bf 36}'$ representation of \SL~\cite{deWit:2002vt} as
\begin{align}
  \Theta _{ab}{}^c{}_d ={\delta^c}_{[a}\theta_{b]d} \,, \qquad
\theta _{ab}=\theta_{(ab)}\,, \qquad
a,b,...=1,...,8 \,,
\label{N8_CP_electric_ET}
\end{align}
the gauge structure constants $X_M$ of ${\rm CSO}(p,q,r)$ are given by
\begin{align}
 X_M =(X_\Lambda , X^\Sigma  ) =(X_{[ab]}, 0) \,, \qquad
X_{[ab]} =\left(
\begin{array}{cc}
X_{[ab][cd]}{}^{[ef]} & 0  \\
0 & X_{[ab]}{}^{[cd]}{}_{[ef]}
\end{array}
\right)\,,
\label{N8_CP_electric_XM}
\end{align}
where
\begin{align}
X_{[ab][cd]}{}^{[ef]}= \delta_{[a}^{~[e}\theta
 _{b][c}\delta_{d]}^{~f]}\,, \qquad
X_{[ab]}{}^{[cd]}{}_{[ef]} =- \delta_{[a}^{~[c}\theta
 _{b][e}\delta_{f]}^{~d]} \,.
\end{align}
In this case the quadratic constraint is automatically satisfied.
Thus any symmetric tensor $\theta _{ab}$ defines
a consistent gauging even if it is noninvertible.

From~(\ref{N8_potential}) and (\ref{N8_CP_electric_ET})
the potential at the extrema is given by
\begin{align}
V_c=\frac 18 g^2 \left[\frac 14 {\rm Tr}(\theta ^2)
-\frac 18 ({\rm Tr} \theta )^2\right]\,.
\label{N8_CP_electric_pot}
\end{align}
In our present normalization, $V_c$ is equivalent to the
cosmological constant.

\subsection{Vacua}

In the electric gauging case, the origin of the
scalar coset corresponds to the critical point
if the following relation holds~\cite{DI}
\begin{align}
2\theta ^2 -\theta {\rm Tr}\theta =2v \mathbb I_{8} \,, \qquad
v:=4 g^{-2}V_c \,.
\label{N8_CP_electricextrema}
\end{align}
Note that the extremum condition is invariant
under $\theta \to P^{-1}\theta P$. Hence we can
confine ourselves to the diagonal $\theta $
by taking $P$ as an orthogonal matrix.
Since the $8\times 8$ matrix $\theta $ obeys a quadratic equation~(\ref{N8_CP_electricextrema}),
its eigenvalues $\lambda_i ~(i=1,2)$ should satisfy
\begin{align}
\lambda^2_i-x \lambda_i -v=0 \,, \qquad x:=\frac 12 {\rm Tr}(\theta ) \,.
\end{align}
Let $n_i $ ($\sum_i n_i=8$) denote the degeneracy of eigenvalue
$\lambda_i$. Then
the extremum condition translates into
\begin{align}
\sum_i(n_i-2) \lambda_i=0\,.
\label{N8_CP_ET_electric_cpcond}
\end{align}
Since we have
\begin{align}
 V_c= -\frac 14 g^2 \lambda_1 \lambda_2\,,
\label{N8_CP_electric_cp_pot}
\end{align}
the potential vanishes for $n_1=2$ or $n_2=2$, in which case
the $\theta $ tensor is noninvertible. One also finds that
the potential is invariant under  $\lambda_i \to -\lambda_i$ and
$\lambda_1\leftrightarrow \lambda_2$, thereby
these cases correspond to the same vacua.

It is observed that
equations~(\ref{N8_CP_electric_pot}) and (\ref{N8_CP_electricextrema})
are invariant under the rescaling $\theta \to e^\alpha \theta $ with
$g \to e^{-\alpha } g~(\alpha \in \mathbb R)$.
Using this freedom,
we are free to set  ${\rm det}\theta =\pm 1$  for $n_i\ne 2$, whereas
for $n_1=2$ ($n_2=2$) we can choose $\lambda_1$ ($\lambda_2$) to take any nonvanishing value.
Under these conditions,
it turns out that the extrema in the electric gaugings are exhausted by
table 1 of reference~\cite{DI}.

\subsection{Mass spectrum}

We now move on to the main part of this paper
and determine analytically the full mass spectrum
of 70 scalars.
Let $s_\rho =(t_\rho +{}^T t_\rho)/2 $ decompose into
\begin{align}
 s_\rho  = \left(
\begin{array}{cc}
s_{\rho [ab]}{}^{[cd]} & s_\rho {}_{[abcd]} \\
s_{\rho}{}^{ [abcd]} & s_\rho {}^{[cd]}{}_{[ab]}
\end{array}
\right) \,,\label{N8_CP_ET_smat}
\end{align}
where
\begin{align}
 s_{\rho [ab]}{}^{[cd]} = - s_\rho {}^{[cd]}{}_{[ab]} =2 (S_\rho
 )_{[a}{}^{[c}{\delta_{b]}}^{d]} \,, \qquad (s_\rho
 )_{[abcd]}=(s_\rho )^{[abcd]} =(U_\rho )_{[abcd]} \,.
\label{N8_CP_ET_scomp}
\end{align}
Each of real tensors ($S, U$) has 35 components and satisfies
\begin{align}
  S={}^T S \,, \qquad {\rm Tr}(S)=0 \,, \qquad U=\star U \,.
\end{align}
Substituting (\ref{N8_CP_electric_XM}) and (\ref{N8_CP_ET_smat})
into (\ref{N8_CP_massmatrix}),
we are led to
\begin{align}
M^2 =M_{(1)}^2 (\theta ) +M_{(2)}^2 (\theta ) \,,
\label{N8_CP_electric_mass}
\end{align}
with
\begin{align}
 M_{(1)}^2(\theta ) &= \frac 18 g^2
\left[-{\rm Tr}(\theta ){\rm Tr}(S^2 \theta ) -[{\rm Tr}(\theta
 S)]^2 +2{\rm Tr}(S^2 \theta ^2 )+2 {\rm Tr}(S\theta S \theta )\right] \,,
\label{N8_CP_electric_M1} \\
M_{(2)}^2(\theta ) &=\frac 18 g^2\left[ -U^2_{[ab][cd]} \theta_{ac}\theta_{bd} +\frac 1{24}U\cdot U {\rm
 Tr} (\theta^2 )\right] \,.
\label{N8_CP_electric_M2}
\end{align}
Here we have introduced the abbreviation
\begin{align}
 U\cdot U =U_{abcd}U_{abcd} \,, \qquad (U^2)_{[ab][cd]}=U_{abef}U_{cdef} \,,
 \qquad (U^2)_{ab}=U_{acde}U_{bcde} =\frac 18 U\cdot U  \delta _{ab} \,,
\end{align}
where the final expression follows from the self-duality of $U$.

We now split the matrix $S$ into $n_1$ and $n_2$ blocks
\begin{align}
 S= \left(
\begin{array}{cc}
 A_{11}& A_{12} \\
{}^T \! A_{12} & A_{22}
\end{array}
\right) \,, \qquad \theta =
 \left(
\begin{array}{cc}
 \lambda_1 \mathbb I_{n_1}&  \\
 & \lambda_2 \mathbb I_{n_2 }
\end{array}
\right) \,,
\label{N8_CP_S_dec_electric}
\end{align}
and define
\begin{align}
 A_{11}=\frac{1}{n_1} {\rm Tr}(A_{11}) \mathbb I_{n_1} +\hat A_{1} \,, \qquad
A_{22}=  -\frac{1}{n_2} {\rm Tr}(A_{11}) \mathbb I_{n_2} +\hat A_{2} \,,
\end{align}
where $\hat A_{1} $ and $\hat A_{2}$ are trace-free parts of
$A_{11}$ and $A_{22} $, respectively.

In order to achieve the correct mass spectrum we
need to canonically normalize the scalar kinetic function.
According to (\ref{N8_CP_SpUSp_indicestr}),
the fluctuations of scalar fields $\delta \phi_{ijkl}$ are given by
\begin{align}
2 S_{[a}{}^{[c}\delta_{b]}{}^{d]}  +{\rm i} U_{abcd}=
\frac 1{16}
\Gamma ^{ij}_{ab}\Gamma^{kl}_{cd}\delta \phi _{ijkl } \,,
\label{N8_CP_scalar_fluctuations}
\end{align}
Then the  scalar kinetic term reads
\begin{align}
\frac{1}{12} \ma P_{\mu ijkl}\ma P^{\mu ijkl}=
\frac{1}{12}|\partial_\mu \phi _{ijkl}|^2 =
\frac{1}{2}\tr ((\partial S)^2)+\frac{1}{12}\partial U\cdot \partial
 U  \,. \label{N8_CP_electric_Skin}
\end{align}
It follows that
\begin{align}
\frac 12 \tr ((\partial S)^2)
=\frac 12 \left[\frac{8}{n_1n_2}(\partial \tr A_{11})^2
+\tr((\partial \hat A_{1})^2)
+\tr((\partial \hat A_{2})^2)+2 \tr (\partial {}^T\!\! A_{12}\partial A_{12})
\right] \,.\label{N8_electric_Skin}
\end{align}
With reference to (\ref{N8_CP_electric_M1}) and (\ref{N8_CP_S_dec_electric}),
the mass matrix $M_{(1)}^2$ can be expressed in terms of
fields ($\tr (A_{11}), \hat A_1, \hat A_2, A_{12}$).
The canonical mass eigenvalues can be read off
in such a way that each coefficient of these fields
agrees with (\ref{N8_electric_Skin}), thereby
\begin{align}
 M_{(1)}^2& =
\frac{8}{n_1n_2}m_{0(\mathbf 1,\mathbf 1)}^2 {\rm Tr}(A_{11})^2
 +m_{1(\mathbf N_1,\mathbf  1)}^2 {\rm Tr}(\hat A_{1}^2) +m_{2(\mathbf 1, \mathbf N_2)}^2
{\rm Tr}(\hat A_{2}^2 )+2 m_{*(\mathbf n_1, \mathbf n_2)}^2 {\rm
 Tr}({}^T\! A_{12}A_{12} )
\,,
\label{N8_CP_M1sq_electric}
\end{align}
where
\begin{align}
 N_1 = \frac 12 (n_1-1)(n_1+2) \,, \qquad
 N_2 = \frac 12 (n_2-1)(n_2+2)\,,
\end{align}
and
\begin{subequations}
\label{N8_CP_ET_m1_electric}
\begin{align}
 m_{0(\mathbf 1,\mathbf 1)}^2 & =\frac{g^2 }{64} [2n_2(2-n_1)\lambda_1^2 +2n_1
 (2-n_2)\lambda_2^2 -(n_1-n_2)^2 \lambda_1 \lambda_2 ]\,, \\
 m_{1(\mathbf N_1,\mathbf 1)}^2 &=\frac{g^2}8 \lambda_1
 [(4-n_1)\lambda_1 -n_2 \lambda_2 ]  \,, \\
m_{2(\mathbf 1, \mathbf N_2)}^2 &=\frac{g^2}8 \lambda_2
 [(4-n_2)\lambda_2 -n_1 \lambda_2 ]  \,, \\
 m_{*(\mathbf n_1, \mathbf n_2)}^2 & =\frac{g^2}{16}(\lambda_1 +\lambda_2 )
[(2-n_1)\lambda_1 +(2-n_2)\lambda_2 ] =0  \,.
\end{align}
\end{subequations}
At the last equality we have used the stationary point
condition~(\ref{N8_CP_ET_electric_cpcond}).
It follows that the $A_{12}$ field is always massless.
Boldface letters in the subscript
denote the representations of ${\rm SO}(n_1)\times {\rm SO}(n_2)$.
This notation manifests multiplicities explicitly, i.e.,
$m^2_{(\mathbf k_1,\mathbf k_2)}$ represents the mass spectrum for fields
with $k_1k_2$ degeneracies.
Note that fluctuations of $\tr (A_{11})$ and
$A_{12}$ exist for $n_1n_2 >0$, while $\hat A_{1}$
($\hat A_{2}$) exists for $n_1>1$ ($n_2>1$).

When $n_1\ne 2, 6$, the cosmological constant is nonvanishing.
So we can normalize the mass spectra
in a unit of the cosmological constant~(\ref{N8_CP_electric_cp_pot}) and
obtain a more comprehensive form
\begin{align}
m_{0(\mathbf 1,\mathbf 1)}^2  =-2 V_{c} \,, \qquad
m_{1(\mathbf N_1,\mathbf 1)}^2  = \frac{4V_c}{n_1-2} \,, \qquad
m_{2(\mathbf 1, \mathbf N_2)}^2  = \frac{4V_c}{n_2-2} \,. \qquad
\label{N8_CP_ET_m1_electric2}
\end{align}
Whereas, for $n_1=2$ we have
\begin{align}
m_{0(\mathbf 1,\mathbf 1)}^2
=m_{2(\mathbf 1, \mathbf{20})}^2=
m_{*(\mathbf 2, \mathbf 6)}^2=0 \,, \qquad
m_{1(\mathbf 2, \mathbf 1)}^2 =\frac 14 g^2 \lambda_1^2 \,.
\end{align}
The $n_1=6$ case can be deduced similarly.

Let us turn to determine
the mass  spectrum of pseudoscalars $U$.
We  decompose the eight indices into
$n_1$ and $n_2 $ blocks,
\begin{align}
 S_1 =\{1, ..., n_1\}\,, \qquad S_2 =\{n_1+1,..., n_1+n_2\}\,.
\end{align}
Let
$\ell $ be a non-negative integer taking values in
the range $0\le \ell \le 4$, $0\le 4-\ell \le n_2$.
Then the basis of antisymmetric four-form is labeled by pairs
$I_1, I_2$, where
$I_1~(I_2)$ is a set of $\ell~(4-\ell)$ indices belonging to $S_1~(S_2)$.
For any four-form $Z_{abcd}$, we find
\begin{align}
 \theta^r{} _{[a}\theta ^s{} _{b}Z_{cd]rs}
=\frac 1{12} \left[\ell (\ell -1)\lambda_1^2
+2 \ell (4-\ell )\lambda_1 \lambda_2 +(4-\ell )(3-\ell )\lambda_2^
 2\right]Z_{abcd} \,,
\end{align}
from which  we are led to
\begin{align}
 (1+\star ) \theta^r{} _{[a}\theta ^s{} _{b}U_{cd]rs}&=2 \mu _\ell
 U_{abcd} \,,
\end{align}
where
\begin{align}
  \mu _\ell = & \frac 1{24}
\left[
\ell (\ell -1)\lambda_1^2
+2 \ell (4-\ell )\lambda_1 \lambda_2 +(4-\ell )(3-\ell )\lambda_2^
 2 +(n_1-\ell) (n_1-\ell -1)\lambda_1^2 \right. \nonumber \\
& \left. \qquad
+2 (n_1-\ell) (4-n_1+\ell )\lambda_1 \lambda_2 +(4-n_1+\ell )(3-n_1+\ell )\lambda_2^
 2
\right] \,.
\end{align}
Then the fluctuation mode of $U$ is labeled by a non-negative integer
$\ell$ satisfying
\begin{align}
 n_1 \le 2 \ell \le {\rm min}(2n_1, 8)\,.
\end{align}
Since the kinetic term of $U$ is given by
$(1/12)\partial U_{abcd}\partial U_{abcd}$,
the normalized mass eigenvalue
$m_{[\ell]}$ reads
\begin{align}
M_{(2)}^2 = \frac{1}{6} m_{[\ell]}^2 U\cdot U \,,
\end{align}
where
\begin{align}
 m_{[\ell] }^2 =&\frac{g^2}{32} \left[
\{n_1-\ell (\ell -1)-(n_1-\ell) (n_1-\ell-1)\}\lambda_1^2
-2 \{\ell (4-\ell)+(n_1-\ell)(4+\ell -n_1)\}\lambda_1\lambda_2
\right.\nonumber \\
& \left. \qquad
+\{n_2-(4-\ell )(3-\ell)-(4+\ell -n_1)(3+\ell -n_1)\}\lambda_2^2
 \right] \,,
\label{N8_CP_ET_m2_electric}
\end{align}
with multiplicities
\begin{align}
 2\ell > n_1 ~~:~~ {}_{n_1} C_\ell \times {}_{8-n_1}
C_{4-\ell } \,, \qquad
2\ell =n_1 ~~:~~ \frac 12   {}_{n_1} C_{n_1/2} \times {}_{8-n_1}
C_{4-n_1/2 }\,.
\end{align}
For $n_1\ne 2, 6$, equation~(\ref{N8_CP_ET_m2_electric})
simplifies in a unit of the cosmological constant to
\begin{align}
  m_{[\ell]}^2 = \frac{2 [2\ell ^2 -2n_1 \ell +(n_1-2) ^2
 ]}{(n_1-6)(n_1-2)} V_c \,.
\label{N8_CP_ET_m2_electric2}
\end{align}
For $n_1 =2$ i.e., $\lambda_2=0$,
we have
\begin{align}
m^2_{[\ell=1](\times 20)}=
m^2_{(\mathbf 2, \mathbf{20})^+}
=\frac 1{16}g^2 \lambda_1^2 \,, \qquad
m^2_{[\ell =2](\times 15)}=m^2_{(\mathbf 1, \mathbf{15})}=0 \,,
\end{align}
where we have denoted multiplicities and representations in the
subscript, and  ``$+$'' stands for the self-duality.
As argued in the next subsection, these massless modes
have nothing to do with the Nambu-Goldstone bosons.

We are now in a position to discuss critical points and mass spectra in the electric
gauging. Our classification exhaustively recovers the list of critical points found by
Dall'Agata-Inverso~\cite{DI},
which we summarize in table~\ref{table:N8_CP_electric}.
Our analytic expressions of mass spectra are in perfect agreement with
the reference~\cite{DI}, in which mass eigenvalues may have been
obtained by diagonalization of $70\times 70 $ mass matrix.  In our method it is
obvious which parts of $\phi_{ijkl}$ belong to scalars ($S$ field)
and pseudoscalars ($U$ field). Moreover, our formulation makes it clear
that the mass spectrum is simply specified
group-theoretically by multiplicities. In particular,
it immediately turns out that
the dS vacua necessarily have unstable mode of $m_0^2=-2 V_c$,
arising from the trace part of $n_1$-block [see~(\ref{N8_CP_ET_m1_electric2})].
For ${\rm SO}(5,3)$ and ${\rm SO(4,4)}$ gaugings , this mode corresponds to the
${\rm SO}(5)\times {\rm SO}(3)$ and ${\rm SO}(4)\times {\rm SO}(4)$
invariant scalar, respectively~\cite{Ahn:2001by,Ahn:2002qga,Kallosh:2001gr}.

With the exception of Minkowski vacua,
we have  normalized the mass eigenvalues by the cosmological constant.
This normalization is  intuitively clear since it measures the
curvature of potential. Then,
inspection of (\ref{N8_CP_ET_m1_electric2})  and
(\ref{N8_CP_ET_m2_electric2}) reveals that  the mass spectrum is
determined
only  by the remaining gauge symmetry at the vacua (i.e., $n_i$ only), rather
than the gauging (actual value of $\lambda$).
This statement will become more persuasive  when we
look into the dyonic case in the subsequent section, where
several vacua of different gaugings can have the same mass spectra.

\begin{table}
\label{table:N8_CP_electric}
\begin{center}
{\small
\begin{tabular}[t]{c||ccccc}
Gauging & $G_{\rm reg} $ &$\Lambda $ &$m^2_S$ & $m^2_U$
& SUSY\\ \hline \hline
 ${\rm SO}(4,4)$  &${\rm SO}(4)\times {\rm SO(4)}$ & dS &
$-2_{(\times 1)}$, $2_{(\times 18)}$, $0_{(\times 16)}$ &
$2_{(\times 18)}, 1_{(\times 16)}, -2_{(\times 1)}$ & none \\
${\rm SO}(5,3)$ &${\rm SO}(5)\times {\rm SO(3)}$ & dS
& $-2_{(\times 1)}$, $\frac 43_{(\times 14)}$, $4_{(\times 5)}$, $0_{(\times 15)}$
 & $2_{(\times 30)}$, $-\tfrac 23_{(\times 5)}$& none\\
${\rm SO}(8)$  &${\rm SO}(7)$ &
AdS & $2_{(\times 1)}$, $-\tfrac 45_{(\times 27)}$, $0_{(\times 7)}$ &
$-\tfrac 25_{(\times 35)}$
& none\\
${\rm SO}(8)$ &${\rm SO(8)}$ & AdS & $-\tfrac 23 _{(\times 35)}$ &
$-\tfrac 23 _{(\times 35)} $ & $N=8$
\\
${\rm CSO}(2,0,6)$ & ${\rm SO}(2)$& Mink. &$0_{(\times 33)}$, $2_{(\times 2)}$   &
$0_{(\times 15)}$, $\tfrac 12_{(\times  20)}$ &
none \\
\hline\hline
\end{tabular}
}
\caption{Mass spectrum for electric gaugings. For the (A)dS vacua,
we normalized by the absolute value of the cosmological constant.
For the ${\rm CSO}(2,0,6)$ gauging, we take a normalization
$g\lambda_1 =2\sqrt 2$. }
\end{center}
\end{table}

\subsection{Spontaneous symmetry breaking}
\label{sec:SSB}


As we have seen above, some scalar fields turn out to be massless.
In this section we discuss the Higgs mechanism for the \SL~electric vacua in order to identify
Nambu-Goldstone directions.

Taking the mixed coset representative~(\ref{N8_CP_mixV}),
we define
\begin{align}
\ma Q_{Mij}{}^{kl} = {\rm i}\Omega^{NP} \ma V_{Nij}X_{MP}{}^Q \ma V_Q{}^{kl}\,,
 \qquad
\ma P_{M ijkl} = {\rm i}\Omega^{NP}\ma V_{Nij}X_{MP}{}^Q \ma V_{Qkl } \,,
\end{align}
where
$\Omega^{MN}=\left(\begin{array}{cc}\mathbb O&\mathbb I\\-\mathbb I&\mathbb  O\end{array}\right)$
is the standard metric of \SP.
These tensors satisfy
\begin{align}
 \ma P_M{}^{ijkl}=\frac{\eta }{24}
\epsilon^{ijklmnpq}\ma P_{M mnpq} \,, \qquad
\ma Q_{Mij}{}^{kl }=\delta_{[i}^{[k} \ma Q_{Mj]}{}^{l]} \,, \qquad \ma
 Q_{M}{}^i{}_j =-\ma Q_{Mj}{}^i \,,\qquad
\ma Q_{M i}{}^i =0 \,.
\end{align}
In terms of the $T$-tensor\footnote{Note that
the prefactor $1/2$ in (\ref{N8_CP_Ttensor_dec})
is necessary to derive (\ref{N8_CP_Ttensor_dec}).
The original paper~\cite{de Wit:2007mt}
seems to have a typo.
}
\begin{align}
 T_{\underline{M}\underline N}{}^{\underline P} :=\frac 12
(\ma V^{-1})_{\underline M}{}^M (\ma V^{-1})_{\underline N}{}^N
\ma V_P{}^{\underline P} X_{MN}{}^P \,,
\label{N8_CP_Ttensor}
\end{align}
$\ma P_M$ and $\ma Q_M$ are expressed as
\begin{align}
T_k{}^{lij} =\frac 34{\rm i}\Omega^{MN}\ma Q_{Mk}{}^l \ma V_N{}^{ij} \,,
 \qquad T_{klmn}{}^{ij}=\frac 12 {\rm i}\Omega^{MN} \ma P_{Mklmn}\ma
 V_N{}^{ij}\,.
\label{N8_CP_Ttensor_dec}
\end{align}
These components of the $T$-tensor are related to the ${\rm SU}(8)$
connections and tensors, as shown below.
Let us introduce the ${\rm SU}(8)$ covariant derivative
by
\begin{align}
 \ma D_\mu \ma V_M{}^{ij}= \partial_\mu \ma V_M{}^{ij}-
\ma Q_{\mu kl}{}^{ij} \ma V_M{}^{kl}
-g A_\mu {}^P X_{PM}{}^N\ma V_N{}^{ij}\,,
\end{align}
where
$\ma Q_\mu $ is an ${\rm SU}(8)$ connection, satisfying
\begin{align}
 \ma Q_{\mu ij}{}^{kl}=\delta_{[i}{}^{[k} \ma Q_{\mu j]}{}^{l]}
\,, \qquad
\ma Q_{\mu i}{}^j =- \ma Q_\mu {}^j{}_i \,, \qquad \ma Q_{\mu i}{}^i =0
 \,.
\end{align}
The ${\rm SU}(8)$ covariant derivative is subjected to the restriction
$\Omega^{MN}\ma V_{Mij}\ma D_\mu \ma V_N{}^{kl}=0$,
giving
\begin{align}
 \ma Q_{\mu i}{}^j  =\frac 23 {\rm i}
(\ma V_{\Lambda ik}\partial_\mu \ma V^{\Lambda jk}-\ma V^\Lambda
 {}_{ik}\partial_\mu \ma V_\Lambda {}^{jk}) -g A_\mu {}^M \ma Q_{Mi}{}^j\,.
\end{align}
Similarly, we can define an ${\rm SU}(8)$
covariant tensor
\begin{align}
\ma P_{\mu ijkl}={\rm i}\Omega^{MN}\ma V_{Mij}\ma D_\mu \ma V_{Nkl} \,,
 \qquad
\ma P_\mu {}^{ijkl} =\frac{\eta }{24}\epsilon^{ijklmnpq}\ma P_{\mu mnpq}\,,
\end{align}
which specifies the scalar kinetic term~(\ref{N8_CP_electric_Skin}).
Then it can be shown that
\begin{align}
 \ma P_{\mu ijkl} ={\rm i}(\ma V_{\Lambda ij}\partial_\mu
 V^\Lambda{}_{kl}-\ma V^\Lambda{}_{ij}\partial_\mu \ma V_{\Lambda kl})-g
A_\mu{}^M \ma P_{Mijkl} \,.
\end{align}

Let us now take a base point $O$ in the moduli space of scalar manifold
and employ the following coset representative
\begin{align}
 \ma V= \ma V (O) \exp \left(
\begin{array}{cc}
 0&\phi \\
\bar \phi & 0
\end{array}
\right)\,, \qquad \phi=(\phi_{ijkl})\,,
\qquad \bar \phi =\star \phi =\eta (\phi^{ijkl})\,.
\end{align}
In this gauge,
we obtain
\begin{align}
 \ma Q_{\mu i}{}^j = -g A_\mu {}^M \ma Q_{Mi}{}^j \,, \qquad
\ma P_{\mu ijkl} =\partial_\mu \phi_{ijkl} -g A_\mu {}^M \ma P_{M ijkl
 }\,.
\end{align}
Under the gauge transformation, $\ma V$ changes as
\begin{align}
 \delta \ma V =-g \Lambda^MX_M\ma V =-2{\rm i} g \Lambda^M \ma V\ma
 V_M{}^{\underline N}T_{\underline N}\,,
\end{align}
which implies
\begin{align}
 \delta \phi_{ijkl} =-g \Lambda^M \ma P_{Mijkl} \,,
\label{N8_CP_Higgs_deltaphi_P}
\end{align}
By the $E_{7(7)}$ isometry,
let $O$ move to the extremum of the potential. It then follows that
the broken gauge symmetry  is described by the condition
$\Lambda^M\ma P_{Mijkl}\ne 0 $.
We find that the corresponding $\delta \phi_{ijkl}$
are the Nambu-Goldstone bosons, which
are responsible for the vector fields to acquire mass.

Let us compute the mass matrix for the
Nambu-Goldstone directions.  We first note that
$\ma P_M$ and $\ma Q_M$ take the following forms
at the origin,
\begin{align}
 \ma P_{[ab]ijkl}=\frac{1}{16}(\Gamma_{ijkl})_{c[a}\theta_{b]c} \,,
 \qquad \ma Q_{[ab]i}{}^j =
\frac{1}{4} \theta_{c[a}(\Gamma_{b]c})^{ij}\,,
\end{align}
where $(\Gamma^{ijkl})_{ab}=\Gamma ^{[ij}_{ac}\Gamma^{kl]}_{cb}$.
Converting back to the ${\rm Sp}(56, \mathbb R)$ representation
via~(\ref{N8_CP_scalar_fluctuations}),
the scalar fluctuation (\ref{N8_CP_Higgs_deltaphi_P}) can be
expressed as
\begin{align}
2 S_{[a}{}^{[c}\delta_{b]}{}^{d]}+{\rm i} U_{abcd}
=\frac{1}{16} \Gamma^{ij}_{ab}\Gamma^{kl}_{cd}\delta \phi_{ijkl }\propto \Gamma^{ij}_{ab}\Gamma^{kl}_{cd}
(\Gamma^{ijkl})_{p[e}\theta _{f]p}\Lambda^{[ef]}
\propto (\Lambda \theta -\theta \Lambda )^{[c}{}_{[a}\delta^{d]}{}_{b]} \,,
\end{align}
where we have used
\begin{align}
 \Gamma^{ij}_{ab}\Gamma^{kl}_{cd}
(\Gamma_{ijkl})_{pq} =128
 (\delta_{p[a}\delta_{b][c}\delta_{d]q}+\delta_{q[a}\delta_{b][c}\delta_{d]p})
+32\delta_{a[c}\delta_{d]b}\delta_{pq} \,.
\end{align}
Therefore, the
Nambu-Goldstone bosons are  encoded only  into the $S$ field of the form,
\begin{align}
 S\propto \Lambda \theta -\theta \Lambda \,, \qquad
\Lambda=-{}^T\!\Lambda \,, \qquad  U=0 \,.
\end{align}
Dividing $\Lambda$ into $n_1$ and $n_2$ blocks
\begin{align}
 \Lambda =\left(
\begin{array}{cc}
\Lambda_{11} & \Lambda_{12} \\
-{}^T\!\Lambda_{12} & \Lambda_{22}
\end{array}
\right) \,,
\end{align}
one can derive
\begin{align}
 S=-(\lambda_1-\lambda_2)\left(
\begin{array}{cc}
\mathbb O_{n_1} & \Lambda_{12}\\
{}^T\!\Lambda_{12} & \mathbb O_{n_2}
\end{array}
\right)\,,
\end{align}
and
\begin{align}
M^2_{(1)}=-\frac 18 g^2 (\lambda_1-\lambda_2)^2(\lambda_1+\lambda_2 )
[(n_1-2)\lambda_1+(n_2-2)\lambda_2 ]\tr ({}^T\!\Lambda_{12}\Lambda_{12}) =0\,,
\end{align}
where we inserted the extremum condition~(\ref{N8_CP_ET_electric_cpcond})
 at the last equality.
Therefore, we conclude that $A_{12}$ field corresponds to the
Nambu-Goldstone bosons of the broken noncompact gauge symmetries.
They would be absorbed as the longitudinal modes of gauge fields
to getting massive.

\section{Dyonic gaugings}
\label{sec:dyonic}

We move on to the case where
additional {\bf 36} charges are turned on,
\begin{align}
 \Theta_{ab}{}^c{}_d ={\delta _{[a}}^c\theta_{b] d} \,, \qquad
\Theta ^{abc}{}_d ={\delta^{[a}}_d\xi^{b]c}\,,
\end{align}
where $\theta $ and $\xi$ are  (possibly noninvertible) symmetric tensors.
Since both electric and magnetic charges are introduced,
we shall refer to it  as dyonic.
The gauge generators are now given by
\begin{align}
 X_{[ab]} =\left(
\begin{array}{cc}
 X_{[ab][cd]}{}^{[ef]}&
0
\\
0 &
X_{[ab]}{}^{[cd]}{}_{[ef]}
\end{array}
\right) \,, \qquad
 X^{[ab]} =\left(
\begin{array}{cc}
 X^{[ab]}{}_{[cd]}{}^{[ef]}&
0
\\
0 &
X^{[ab][cd]}{}_{[ef]}
\end{array}
\right) \,,
\label{N8_CP_dyonic_Xtensors}
\end{align}
where
\begin{align}
  X_{[ab][cd]}{}^{[ef]}& =\delta_{[a}{}^{[e}\theta_{b][c}\delta_{d]}{}^{f]}
\,, \qquad
X_{[ab]}{}^{[cd]}{}_{[ef]}
=-\delta_{[a}{}^{[c}\theta_{b][e}{\delta_{f]}}^{d]}
\nonumber \\
 X^{[ab]}{}_{[cd]}{}^{[ef]}&=
-{\delta_{[c} }^{[a}\xi^{b][e}\delta_{d]}{}^{f]} \,, \qquad
{X^{[ab][cd]}}_{[ef]}={\delta_{[e} }^{[a}\xi^{b][c}{\delta_{f]}
 }^{d]}\,.
\end{align}
The value of the potential at the origin
gives the cosmological constant,
\begin{align}
 V_c&=\frac{g^2}{8}\left[\frac 1 4 {\rm Tr}(\theta ^2 )-\frac 1{8} {\rm Tr}(\theta )^2
 +\frac 14 {\rm Tr}(\xi ^2 )-\frac 1{8} {\rm Tr}(\xi )^2\right]
\,.\label{N8_CP_mag_pot2}
\end{align}

\subsection{Vacua}

The extremum condition boils down to~\cite{DI}
\begin{align}
2(\theta ^2 -\xi^2 ) -(\theta {\rm Tr}\theta -\xi {\rm Tr}\xi ) =2 a
 \mathbb I_{8} \,,
\label{N8_CP_spcond1}
\end{align}
where
$a$ is an arbitrary  real constant.
The solution for the quadratic constraint is given by
\begin{align}
\xi =c \theta ^{-1}  \quad (c \in \mathbb R) \,, \qquad
{\rm or}\qquad \xi \theta =0 \,.
\end{align}
These cases will be discussed separately in the following.

\bigskip\noindent{(I) $\theta \propto \xi^{-1}$.}
We start with the discussion for the case in which both
$\theta $ and $\xi $ are invertible.
Letting
\begin{align}
 x :=\frac 12 {\rm Tr}(\theta ) \,, \qquad y:=\frac 12 {\rm Tr}(\theta
 ^{-1}) \,,
\end{align}
the stationary point condition (\ref{N8_CP_spcond1})
can be equivalently written as
\begin{align}
 \theta ^4 -x \theta ^3 -a \theta ^2 +c^2 y \theta -c^2 \mathbb I_{8} =0\,.
\label{N8_CP_spcond2}
\end{align}
Since equation~(\ref{N8_CP_spcond2}) is invariant
under the similarity transformation $\theta \to P\theta P^{-1}$,
we can restrict to diagonal $\theta $. Moreover  equations~(\ref{N8_CP_mag_pot2}) and (\ref{N8_CP_spcond2})
  are invariant under the
rescaling $\theta \to e^\alpha \theta $
with $c\to e^{2\alpha} c,  g\to e^{-\alpha} g$
($\alpha \in \mathbb R$).
Noticing that  the embedding tensor
arises together with the coupling constant,
we can achieve $c=1 $ without loss of generality.

Since $\theta $ obeys a quartic polynomial,
it has four eigenvalues $\lambda_i$ ($i=1,...,4$) with
degeneracy $n_i (\ge 0)$,
\begin{align}
\theta =\lambda_1 \mathbb I_{n_1}\oplus
\lambda_2 \mathbb I_{n_2}\oplus
\lambda_3 \mathbb I_{n_3}\oplus
\lambda_4 \mathbb I_{n_4}\,, \qquad
 \sum_i n_i=8 \,.
\end{align}
From (\ref{N8_CP_spcond2}) one can easily derive
\begin{align}
x=\sum_i \lambda_i \,, \qquad  y
 =-\sum_{i<j<k}\lambda_i\lambda_j\lambda_k \,,
\qquad a =-\sum_{i<j}\lambda_i \lambda_j \,, \qquad
\prod_i\lambda_i =-1\,.
\label{N8_CP_xycond}
\end{align}
Hence $\lambda_i$'s satisfying the following relation correspond to
the critical point,
\begin{align}
\sum_i (n_i-2) \lambda_i =0 \,, \qquad \sum_i \frac{n_i-2}{\lambda_i}=0
 \,, \qquad
\prod _i \lambda_i =-1 \label{N8_CP_xycond2}
\end{align}
Substitution of~(\ref{N8_CP_xycond2}) into~(\ref{N8_CP_mag_pot2})
yields
\begin{align}
 V_c = \frac{g^2}{32}\sum_i (n_i-2)(\lambda _i^2 +\lambda_i^{-2}) \,.
\label{N8_CP_mag_pot3}
\end{align}
Therefore neither the ordering of $\lambda_i$ nor the overall sign flip
$\lambda_i \to -\lambda_i$  affect the scalar potential.

We are now going to classify all critical points
satisfying~(\ref{N8_CP_xycond2}).
Letting us denote $p_i:=n_i-2$,
$\sum_i p_i =0$ and $-2 \le p_i \le 6$ must be satisfied. Hence there are 15
possible combinations of $\{p_i\}$, which can be
categorized into the following 3 groups,
\begin{align}
({\rm i}):&\{(4,-2,-2,0), (3,-2,-1,0), (2,-2,0,0),
(2,-1,-1,0), (1,1,-2,0), (1,-1,0,0), (0,0,0,0) \},\nonumber \\
({\rm ii}):&\{
(2,-2,1,-1), (1,-1,1,-1), (2,-2,2,-2)\}\,,  \nonumber \\
({\rm iii}): &\{
(6,-2,-2,-2), (5,-1,-2,-2), (3, 1, -2,-2),
(4,-2, -1,-1), (3,-1,-1,-1) \} \nonumber\,.
\end{align}
Since the ordering of $p_i$'s is irrelevant,
we can take
(i)~$p_4=0$, (ii)~$p_1=-p_2$ with $p_3=-p_4$ and
(iii)~$p_3=p_4$, respectively without losing generality.
In the following,
we shall discuss separately these cases.

\bigskip\noindent{(i) $p_4=0$.}
Equation~(\ref{N8_CP_xycond2}) implies
that all cases belonging to this family
can be identified as degenerate cases of $p_i=0$ ($i=1,...,4$).
Hence the $\theta $ tensor can be written as
\begin{align}
 \theta =r \mathbb I_2 \oplus s \mathbb I_2
\oplus t \mathbb I_2 \oplus\left(-\frac{1}{rst}\right)\mathbb I_2 \,,
\qquad {\rm det} \theta =1 \,,
\label{N8_CP_SO62_theta}
\end{align}
where $r$, $s$ and $t$ are real parameters.
We find that the  cosmological constant (\ref{N8_CP_mag_pot3}) vanishes
and one of the eigenvalues must have opposite sign from others,
since the overall sign flip has no effect.
Hence these vacua correspond to the ${\rm SO}(6,2 )$ gauging,
which spontaneously breaks down to a  compact group
${\rm SO}(2)\times{\rm SO}(2)\times{\rm SO}(2)\times{\rm SO}(2)$
at the vacua.
The residual gauge symmetry would
be enhanced to ${\rm SO}(4)\times{\rm SO}(2)\times{\rm SO}(2)$
for $s=r$ and to ${\rm SO}(6)\times{\rm SO}(2)$ for $r=s=t$.

As we have seen, these vacua are parametrized by 3 continuous parameters.
It is noted that
the determinant remains
invariant ${\rm det}\ti \theta ={\rm det}\theta $
under the \SL~transformation
$\theta \to \ti \theta = {}^T\!U\theta U$
(${\rm det}U=1$).  If ${\rm det}\theta =\pm 1$
had not been satisfied,
it would correspond to the deformation of the theory.
This is not the case now,
since ${\rm det}\theta =1$ is always satisfied.
This is consistent with the fact that the moduli mass matrix
vanishes exactly in the directions corresponding to the variation
of these parameters, as we will see later.

\bigskip\noindent{(ii) $p_1=-p_2$ and $p_3 =-p_4(p_i\ne 0)$.}
In this case,
$(n_1,n_2,n_3,n_4)=(4,0,3,1), (3,1,3,1), (4,0,4,0)$
are relevant.
Inserting $\lambda _4 =-1/(\lambda_1\lambda_2\lambda_3)$
into the first two equations of (\ref{N8_CP_xycond2}),
we get two quadratic equations for $\lambda_3$,
\begin{align}
 p_3 \lambda_1 \lambda_2 \lambda_3^2+p_1 \lambda_1\lambda_2 (\lambda_1
 -\lambda_2 ) \lambda_3
+p_3 =0 \,,\qquad
p_3 \lambda_1\lambda_2 \lambda_3^2 +p_1 (\lambda_1^{-1}
-\lambda_2^{-1}) \lambda_3 +p_3=0\,.
\label{N8_CP_II_quadeq}
\end{align}
These equations imply
$[(\lambda_1\lambda_2)^2+1](\lambda_1 -\lambda_2)=0$, giving
$\lambda _2=\lambda_1$, $\pm {\rm i}\lambda_1$.
In
the $\lambda_2=\lambda_1$ case, equation~(\ref{N8_CP_II_quadeq})
implies that $\lambda_i$ cannot be all real, so that only the
$(n_1,n_2,n_3,n_4)=(4,0,4,0)$ case is possible.
The $\lambda_2 =\pm {\rm i}/\lambda_1$ case
amounts to the permutations of eigenvalues for the
$\lambda_2=\lambda_1$ case.
Hence,
the $(4,0,3,1), (3,1,3,1)$ types have no fixed points and
this class of solution corresponds to the
${\rm SO}(4, 4)$ dS vacua,
\begin{align}
 \theta =\lambda \mathbb I_4 \oplus (-\lambda) \mathbb I_4 \,, \qquad
V=\frac{g^2}{4} \left(\lambda^2 +\lambda^{-2}\right) \,, \qquad
{\rm det}\theta =\lambda^8>0 \,, \qquad \lambda \in \mathbb R\,.
\end{align}
At the vacua, the noncompact gauge symmetry is spontaneously broken
to ${\rm SO}(4)\times {\rm SO}(4)$.

Up to this point, we are left with a single parameter $\lambda $.
Usually we set ${\rm det }\theta =\pm 1$, giving
$\lambda=1$ and $V=g^2/2$. Previous studies which did not employ the
embedding tensor formalism have imposed this relation, so
any particular attention has been paid to this freedom.
However, it appears that this remaining freedom implies that
we have a one-parameter family of ${\rm SO}(4,4 )$ deformed theories.
This is in sharp contrast with the case (i), for which
${\rm det}\theta =1$ is always fulfilled.
Now ${\rm det }\theta =\pm 1$ is not satisfied,
hence it cannot be transformed by the  \SL~action
to ${\rm det}\ti \theta =\pm 1$,
implying the deformation of the theory.

Although
it is important to show which parameter region
corresponds to the equivalent theories,
this issue is in general difficult and
beyond the scope of the present article.\footnote{Recently
it has been conjectured that
the different theories may be distinguished according to the
eigenvalues of tensor classifier constructed from a
quartic invariant of $E_{7(7)}$,  in analogy with the
black hole geometry~\cite{Dall'Agata:2012bb}. }
Hence, we will simply specify the allowed range of deformation
parameter.
However, as far as the stability issue is concerned, the mass spectrum is
nevertheless insensitive to the deformation parameter as we will prove in the next subsection.

\bigskip\noindent{(iii) $p_3 =p_4(\ne 0)$.}
We next discuss the $p_3=p_4$ case, viz,
$(n_1,n_2,n_3,n_4)=(8,0,0,0)$,
$(7,1,0,0)$, $(5,3,0,0)$, $(6,0, 1,1)$ and $(5,1,1,1)$.
Inserting $\lambda _4 =-1/(\lambda_1\lambda_2\lambda_3)$
into the first two equations of (\ref{N8_CP_xycond2}),
we get two quadratic equations for $\lambda_3$,
\begin{align}
 p_3 \lambda_1\lambda_2 \lambda_3^2 +\lambda_1\lambda_2
 (p_1\lambda_1+p_2 \lambda_2 )\lambda_3 -p_3 =&0 \,,\qquad
p_3 \lambda_1 \lambda_2 \lambda_3 ^2 -
 \left(\frac{p_1}{\lambda_1}+\frac{p_2}{\lambda_2}
\right) \lambda_3 -p_3 =0 \,,
\label{N8_CP_genlam3_quadeq}
\end{align}
Compatibility of these equations translates into
a cubic equation for $\lambda_2$,
\begin{align}
p_2 \lambda_1^2 \lambda_2 ^3 +p_1 \lambda_1^3 \lambda_2 ^2 +
p_1 \lambda_2 + p_2 \lambda_1 =0 \,.
\label{N8_CP_genlam_cubiceq}
\end{align}
The solution of the above equation
can be most conveniently parametrized as
\begin{align}
\lambda_1 = \sqrt{-\frac{s(p_2 s^2+ p_1)}{p_1 s^2 + p_2 }} \,, \qquad
\lambda_2 =\frac{s}{\lambda_1 } \,,
\label{N8_CP_lambda_s_sol}
\end{align}
where $s (\ne 0, \pm \sqrt{- p_2/p_1}, \pm \sqrt{-p_1/p_2})$
is a real parameter (it leads to the contradiction
if $s$ is complex). In this case, the cosmological constant~(\ref{N8_CP_mag_pot3})
reduces to
\begin{align}
V_c =- \frac{g^2 p_1p_2 (p_1+p_2)(1+s^2)^3}{16 s (p_1s^2+p_2)
 (p_2s^2+p_1)} \,.
\label{n8_CP_dyonic_III_pot}
\end{align}
We now take a closer look at each vacuum.

\bigskip\noindent{(8,0,0,0):}
The $\theta $ tensor and the potential are given by
\begin{align}
 \theta =\lambda \mathbb I_{8} \,, \qquad V_c=
 -\frac{3g^2(1+\lambda^4)}{4\lambda^2 }\,.
\label{N8_CP_dyonic_theta_8080}
\end{align}
$\lambda \in \mathbb R$ is a deformation parameter.
If we require ${\rm det}\theta =1$, we have $\lambda=1$ as usual.
This is the well-known maximally supersymmetric AdS vacua
at which all (pseudo)scalars vanish.

\bigskip\noindent{(7,1,0,0):}
The $\theta $ tensor and the potential are given by
\begin{align}
 \theta & =\lambda \mathbb I_7 \oplus \frac{s}{\lambda }\mathbb I_{1} \,, \qquad
\lambda = \sqrt{\frac{s(s^2-5)}{5s^2-1}} \,, \nonumber \\
V&=
 -\frac{5g^2 (1+s^2)^3}{4s(-5+s^2)(-1+5s^2)}\,, \qquad
{\rm det}\theta =\frac{s^4(-5+s^2)^3}{(-1+5s^2)^3} \,.
\label{N8_CP_dyonic_theta_7100}
\end{align}
$s $ is a deformation parameter.
When $0<s<1/\sqrt 5$, $\sqrt 5<s$, we have
${\rm det}\theta>0$, producing AdS critical points of the  ${\rm SO}(8)$ gauging.
Whereas,  the parameter region $-\sqrt 5 <s<-1/\sqrt 5$ (${\rm det}\theta <0$)
provides  AdS critical points of the  ${\rm SO}(7,1)$ gauging.
If we require ${\rm det}\theta =\pm 1$,
the former case gives
$V_c=-25\sqrt 5 g^2/32$ with  $s=\sqrt 5 \pm 2$,
and the latter case  gives $V_c=-5g^2/8$ with $s=-1$.
At these vacua, the gauge symmetries are
spontaneously broken to ${\rm SO}(7)$.

\bigskip\noindent{(5,3,0,0):}
The $\theta $ tensor and the potential are given by
\begin{align}
 \theta &=\lambda \mathbb I_5 \oplus \frac{s}{\lambda }\mathbb I_{3} \,, \qquad
\lambda = \sqrt{-\frac{s(3+s^2)}{3s^2+1}} \,, \nonumber \\
V&=
 -\frac{3g^2 (1+s^2)^3}{4s(3+s^2)(1+3s^2)}\,, \qquad
{\rm det}\theta =-\frac{s^4(3+s^2)}{1+3s^2}<0 \,.
\label{N8_CP_dyonic_theta_5300}
\end{align}
$s(<0)$  is a deformation parameter.
This case yields the  dS vacua  of the  ${\rm SO}(5,3)$ gauging
 with a residual
gauge symmetry ${\rm SO}(5)\times {\rm SO}(3)$.
Assuming ${\rm det}\theta =-1$, we have
$V_c=3g^2/8$ with $s=-1$.

\bigskip\noindent{(6,0,1,1):}
We have
\begin{align}
 \theta &=\lambda \mathbb I_6 \oplus \lambda_+ \mathbb I_1 \oplus
 \lambda_- \mathbb I_{1} \,, \qquad
\lambda =\sqrt{\frac{s(s^2-2)}{2s^2-1}} \,, \qquad
\lambda _\pm = \pm \sigma  \frac{s(\sqrt 2 s\mp 3)+\sqrt 2}{\sqrt{s (2s^2-1)(s^2-2)}}
\nonumber \\
V&=
 -\frac{g^2(1+s^2)^3}{4s(2s^2-1)(s^2-2)}\,, \qquad
{\rm det}\theta =-\frac{s^2(-2+s^2)^3}{(-1+2s^2)^3} \,,
 \label{N8_CP_dyonic_theta_6100}
\end{align}
where $s$ is a deformation parameter,
$\sigma =-1$ for
$0<s<1/\sqrt 2$ and $\sigma=1$ otherwise.
For $-\sqrt 2<s<-(1/\sqrt 2)$, we have ${\rm det}\theta >0$, whereby
AdS vacua  of the  ${\rm SO}(8)$ gauging are realized.
For $s>\sqrt 2$ and $0<s<1/\sqrt 2$, we have ${\rm det}\theta<0$,
corresponding to the AdS vacua of the   ${\rm SO}(7,1)$ gauging.
In both cases the remaining gauge symmetry is
${\rm SO}(6)$.
If we suppose ${\rm det}\theta =\pm 1$,
${\rm det} \theta =1$ gives
$V_c=-2 g^2$ at $s=-1$, while
${\rm det} \theta =-1$ gives
$V_c\sim -0.6896 g^2$  at
$s= [(7+3\sqrt 3\pm\sqrt{72+42\sqrt 3})/2]^{1/2}$.
The last vacua seem new ones in the
undeformed ${\rm SO}(7,1)$ gauged supergravity.

\bigskip\noindent{(5,1,1,1):}
We have
\begin{align}
 \theta =&\lambda \mathbb I_5 \oplus \frac{s}{\lambda} \oplus
\lambda_+ \mathbb I_1 \oplus \lambda_- \mathbb I_{1} \,, \qquad
\lambda = \sqrt{\frac{s(s^2-3)}{3 s^2-1}}\,,\qquad
\lambda_\pm = \pm \sigma \frac{s(\sqrt 3 s \mp 4)+\sqrt 3}{\sqrt{s (s^2-3)(3
 s^2 -1)}}\,,
\nonumber \\
V=&
 -\frac{3g^2(1+s^2)^3}{8s (3s^2-1)(s^2-3)}\,, \qquad
{\rm det}\theta =-\frac{s^2(-3+s^2)^2}{(1-3s^2)^2}<0 \,.
\end{align}
$s(\ne 0, \pm 1/\sqrt 3, \pm \sqrt 3)$
is a deformation parameter,
$\sigma =-1 $ for $0<s<1/\sqrt 3$ and $\sigma=1$
otherwise.
This gives AdS critical points
of the ${\rm SO}(7,1)$ gauging.
When ${\rm det}\theta =-1$ we have
$V_c=-3g^2/4$ at $s=-1, 2\pm \sqrt 3$.
However it turns out that these
cases simply relabel the eigenvalues hence give the
equivalent vacua.

\bigskip\noindent{(II) $\theta \xi=0$. }
We shall next discuss the
$\theta \xi=0$ case, in which $\theta $ and $\xi $
can be taken to be
\begin{align}
 \theta =\ti \theta\oplus \mathbb O_{n_3+n_4} \,, \qquad
 \xi =\mathbb O_{n_1+n_2}\oplus \ti \xi \,,  \qquad
\sum_i n_i =8 \,,
\end{align}
where $\ti \theta $ and $\ti \xi $ are
$(n_1+n_2)\times (n_1+n_2)$ and $(n_3+n_4)\times (n_3+n_4)$
matrices, respectively. These tensors
give decoupled quadratic equations,
\begin{align}
2 \ti \theta ^2 -\ti \theta \tr (\ti \theta )=2 a \mathbb I_{n_1+n_2}  \,, \qquad
-2 \ti \xi ^2 +\ti \xi  \tr (\ti \xi )=2 a\mathbb I_{n_2+n_4}  \,,
\label{N8_CP_zero_CPeq}
\end{align}
where $a $ is a constant.
So they can be simultaneously taken to be
diagonal forms,
\begin{align}
\theta = \left(
\begin{array}{ccc}
\lambda_1 \mathbb I_{n_1}&  &  \\
& \lambda_2 \mathbb I_{n_2} & \\
& & \mathbb O_{n_3+n_4}
\end{array}
\right) \,, \qquad
\xi = \left(
\begin{array}{ccc}
\mathbb O_{n_1+n_2}&  &  \\
&   \kappa _3 \mathbb I_{n_3}& \\
& & \kappa _4 \mathbb I_{ n_4}
\end{array}
\right) \,.
\end{align}
The critical point condition~(\ref{N8_CP_zero_CPeq}) reduces to
\begin{align}
\sum_{i=1}^2 (n_i-2)\lambda_i =0 \,, \qquad
\sum _{i=3}^4( n_i-2)  \kappa_i =0 \,, \qquad \lambda_1\lambda_2 =-
 \kappa_3 \kappa_4 \,.
\label{N8_CP_dyonic_deg_CPcond}
\end{align}
Using this condition,
the potential is now given by
\begin{align}
 V_c =\frac{g^2}{32} \left[\sum_{i=1}^2 (n_i-2)\lambda_i^2
+\sum_{i=3}^4 (n_i-2) \kappa_i^2 \right] \,.
\label{N8_CP_pot_deg}
\end{align}
Since~(\ref{N8_CP_zero_CPeq}) is invariant
under $\theta \to e^\alpha \theta $, $\xi \to e^\alpha \xi$ with
$a\to e^{2\alpha }a$ ($\alpha \in \mathbb R$),
only one of the eigenvalues can take any value we wish,
since  $\theta $ and $\xi $
cannot be rescaled independently.
Note also the invariance under $\theta \to -\theta $ and $\xi\to -\xi$,
and $\theta \leftrightarrow \xi $.
We will below examine the vacuum structure and the mass spectrum
depending on if $n_i-2$ is zero or not.

\bigskip\noindent{(i) $n_4=2$.}
In this case we can infer that the potential vanishes and the Minkowski vacua are
realized.
Taking the rescaling freedom into account,
two kinds of gaugings are possible.

\begin{itemize}
\item[($a$)]
One is the ${\rm SO}(4)\times {\rm SO}(2,2)\ltimes \mathbb T^{16}$
gauging, for which
\begin{align}
 \theta =\left(
\begin{array}{ccc}
s \mathbb I_{2} & & \\
& (1/s) \mathbb I_{2} & \\
& & \mathbb O_4
\end{array}
\right)\,, \qquad
\xi = \left(
\begin{array}{ccc}
\mathbb O_4 & & \\
& t \mathbb I_2 & \\
&& -(1/t) \mathbb I_2
\end{array}
\right) \,,
\label{N8_CP_mag_deg_thetaxi_case1}
\end{align}
where $s$ and $t$ are real parameters.
At the vacua the gauge symmetry is broken to
${\rm SO}(2)\times{\rm SO}(2)\times{\rm SO}(2)\times{\rm SO}(2)$.

\item[($b$)]
The other is the  ${\rm SO}(2)\times {\rm SO}(2)\ltimes \mathbb T^{20}$
gauging, for which
\begin{align}
  \theta =\left(
\begin{array}{cc}
 \mathbb I_{2}  & \\
 & \mathbb O_6
\end{array}
\right)\,, \qquad
\xi = \left(
\begin{array}{cc}
\mathbb O_6 &  \\
& s \mathbb I_2
\end{array}
\right) \,,
\label{N8_CP_mag_deg_thetaxi_case2}
\end{align}
where $s$ is a real parameter.
The residual gauge symmetry is
${\rm SO}(2)\times {\rm SO}(2)$.
\end{itemize}

\bigskip\noindent{(ii) $n_i \ne 2$.}
Assuming $n_i\ne 2$, we can obtain
\begin{align}
\lambda_2 =-\frac{n_1-2}{n_2 -2} \lambda_1 \,, \qquad
\kappa_3 = \sqrt{
-\frac{(n_1-2)(n_4-2)}{(n_2-2)( n_3-2)}
} \lambda_1\,, \qquad  \kappa_4 =-\frac{n_3-2}{ n_4 -2} \kappa_3
 \,.
\end{align}
The possible values of $n_i(\ne 2)$ and the corresponding
gaugings are given by
\begin{subequations}
\begin{align}
(7,0,1,0)~&:~{\rm SO}(7) \ltimes \mathbb T^7  \,, \\
(6,1,0,1)~&:~{\rm SO}(7) \ltimes \mathbb T^7  \,, \\
(6,0,1,1)~&:~ {\rm SO}(6)\times {\rm SO}(1,1)\ltimes \mathbb T^{12} \,, \\
(5,1,1,1) ~&:~ {\rm SO}(6)\times {\rm SO}(1,1)\ltimes \mathbb T^{12} \,.
\end{align}
\end{subequations}
Otherwise, eigenvalues will be imaginary.
The potential now reads
\begin{align}
V_c=\frac{g^2(n_1-2)}{16(n_2-2)}(n_1+n_2-4)\lambda _1^2 \,.
\end{align}
One can easily verify that all vacua falling into this family correspond to AdS.
Eliminating $\lambda_1$ by the rescaling freedom,
no tunable parameters are left and
we arrive at the following exhaustive list.

\bigskip\noindent{(7,0,1,0):}
We have ${\rm SO}(7) \ltimes \mathbb T^7$ gauging with
\begin{align}
 \theta =\mathbb I_7 \oplus  0 \,, \qquad
\xi = \mathbb O_7 \oplus (\sqrt 5) \,, \qquad
V_c= -\frac{15}{32}g^2 \,.
\label{N8_CP_dyonic_deg_theta_7010}
\end{align}
The gauge symmetry is broken to ${\rm SO}(7)$.

\bigskip\noindent{(6,1,0,1):}
We have ${\rm SO}(7) \ltimes \mathbb T^7$ gauging with
\begin{align}
 \theta =\mathbb I_6 \oplus 4 \oplus 0 \,, \qquad
 \xi = \mathbb O_7 \oplus (-2\sqrt 2)\,, \qquad
V_c=-\frac{3}{4}g^2  \,.
\label{N8_CP_dyonic_deg_theta_6110}
\end{align}
The gauge symmetry is broken to ${\rm SO}(6)$.

\bigskip\noindent{(6,0,1,1):}
We have ${\rm SO}(6)\times {\rm SO}(1,1)\ltimes \mathbb T^{12}$ gauging with
\begin{align}
 \theta =\mathbb I_6 \oplus \mathbb O_2 \,, \qquad
 \xi = \mathbb O_6  \oplus(\sqrt 2)\oplus (-\sqrt 2) \,, \qquad
V_c=-\frac{1}{4}g^2  \,. \label{N8_CP_dyonic_deg_theta_6011}
\end{align}
The gauge symmetry is broken to ${\rm SO}(6)$.

\bigskip\noindent{(5,1,1,1):}
We have ${\rm SO}(6)\times {\rm SO}(1,1)\ltimes \mathbb T^{12}$ gauging with
\begin{align}
\theta =\mathbb I_5 \oplus (3) \oplus \mathbb O_2 \,, \qquad
\xi = \mathbb O_6 \oplus (\sqrt 3) \oplus (-\sqrt 3)
 \,, \qquad
V_c= -\frac{3 }{8}g^2 \,.
\label{N8_CP_dyonic_deg_theta_5111}
\end{align}
The gauge symmetry is broken to ${\rm SO}(5)$.

\subsection{Mass spectrum}

Inserting (\ref{N8_CP_ET_smat})
and (\ref{N8_CP_dyonic_Xtensors}) into~(\ref{N8_CP_massmatrix}), we can show
after a rather lengthy but straightforward computation
that the mass spectra for $S$ and $U$ are the
simple sum of $\theta $ and $\xi $ terms,
\begin{align}
M_{(1)}^2 =M_{(1)}^2 (\theta )+M_{(1)}^2 (\xi ) \,,
\qquad
M_{(2)}^2 =M_{(2)}^2 (\theta )+M_{(2)}^2 (\xi ) \,,
\label{N8_CP_dyonic_massformula}
\end{align}
where $M_{(i)}(\theta )$ is given by
(\ref{N8_CP_electric_M1}) and (\ref{N8_CP_electric_M2}).
When $\theta $ has a structure of $n_1$ and $n_2$ blocks only
(i.e., $n_3=n_4=0$),
it turns out that equations
(\ref{N8_CP_ET_m1_electric2}) and
(\ref{N8_CP_ET_m2_electric2}) continue to hold in the
dyonic case because $M_{(i)}^2$ and
$V_c$  enjoy the property that $\theta $ and
$\xi$ terms are decoupled, so that
they sum up to give the same contributions.

We give general formulas applying to
all dyonic cases.
Following the same argument in the preceding section,
we decompose $\theta $ and $\xi $ into $n_1+n_2+n_3+n_4$ blocks,
\begin{align}
\theta =\left(
\begin{array}{cccc}
\lambda_1 \mathbb I_{n_1} & & & \\
& \lambda_2 \mathbb I_{n_2}& & \\
& & \lambda_3 \mathbb I_{n_3}& \\
& & & \lambda_4 \mathbb I_{n_4}
\end{array}
\right) \,,
\qquad
\xi = \left(
\begin{array}{cccc}
\kappa _1 \mathbb I_{n_1} & & & \\
& \kappa_2 \mathbb I_{n_2}& & \\
& & \kappa_3 \mathbb I_{n_3}& \\
& & & \kappa_4 \mathbb I_{n_4}
\end{array}
\right) \,.
\label{N8_CP_Stheta_4block}
\end{align}
and correspondingly
\begin{align}
S=\left(
\begin{array}{cccc}
A_{11} & A_{12}& A_{13}& A_{14} \\
A_{21} &A_{22} & A_{23} & A_{24} \\
A_{31} & A_{32} & A_{33} & A_{34} \\
A_{41} & A_{42} & A_{43} &A_{44}
\end{array}
\right)\,,
\end{align}
where $A_{ij}$ is an $n_i \times n_j$ matrix
satisfying
\begin{align}
A_{ij}=\,{}^T\!A_{ji} \,, \qquad
\sum_i n_i A_{ii}=0 \,.
\end{align}
Denoting
$A_{ii}=\hat A_i + (1/n_i){\rm Tr}(A_{ii})$,
the mass matrix for $S$ is given by
\begin{align}
M_{(1)}^2 = \sum_{i,j}\mu_{ij}^2\tr (A_{ii}) \tr (A_{jj}) +\sum_i
m_i^2  {\rm Tr} (\hat A_i^2)+2\sum_{i<j} m_{ij*}^2 \tr ({}^T\!A_{ij}A_{ij}) \,,
\label{N8_CP_dyonic_M1sq_dec}
\end{align}
with
\begin{subequations}
\begin{align}
\mu_{ij}^2
 &=\frac{g^2}{8}\left[\frac{1}{n_i}\left\{
4(\lambda_i^2+\kappa_i^2)-\lambda_i
 \sum_k n_k\lambda_k -\kappa _i\sum_kn_k\kappa_k  \right\}\delta_{ij}-(\lambda_i\lambda_j
 +\kappa_i\kappa_j )\right] \,, \\
m_i^2 &=\frac{g^2}{8} \left[4 (\lambda_i^2+\kappa_i^2)
-\lambda_i \sum_jn_j\lambda_j- \kappa_i \sum_j n_j\kappa_ j
\right] \,, \\
m_{ij*}^2&=\frac{g^2}{8}\left[-\frac 12(\lambda_i+\lambda_j)\sum_kn_k\lambda_k
+ (\lambda_i+\lambda_j)^2-\frac 12 (\kappa_i+\kappa_j)\sum_kn_k\kappa_k
+ (\kappa _i+\kappa _j)^2 \right]=0\,.
\end{align}
\end{subequations}
In the last step, we have used the
critical point conditions (\ref{N8_CP_xycond2})
and (\ref{N8_CP_dyonic_deg_CPcond}).
Hence $A_{ij}~(i<j)$ field is always massless.
Repeating the parallel argument in section~\ref{sec:SSB},
we can find that this direction corresponds to the
Nambu-Goldstone bosons.

In order to achieve the canonical normalization of $S$,  we
have to eliminate $\tr (A_{44})$ by the condition $\tr (S)=0$.
After some simple algebra, we obtain the diagonal form,
\begin{align}
\frac 12 (\partial \tr(S))^2 =\frac 12 \left[
\sum_{i=1}^4
(\partial \tr (\hat A_i))^2
+2 \sum_{i<j}\tr (\partial
A_{ij}\partial {}^T\! A_{ij} )
+\sum_{i=1}^3
(\partial  a_i)^2
\right]  \,,
\label{N8_CP_dyonic_Sdec_kin}
\end{align}
where $a_i$'s are canonically normalized scalars and defined by
\begin{subequations}
\label{N8_CP_dyonic_canscalars_tr}
\begin{align}
 a_1:= & \sqrt{\frac{n_1+n_2+n_3+n_4}{n_1(n_2+n_3+n_4)}} \tr(A_{11}) \,, \\
 a_2:=& \sqrt{\frac{n_2+n_3+n_4}{n_2 (n_3+n_4)}} \left[\tr(A_{22})+\frac{n_2
}{n_2+n_3+n_4} \tr(A_{11})\right] \,, \\
 a_3:=& \sqrt{\frac{n_3+n_4}{n_3n_4}}\left[\tr(A_{33})
 +\frac{n_3}{n_3+n_4}(\tr(A_{11})+\tr(A_{22}))\right] \,.
\end{align}
\end{subequations}
Note that $a_3$ should be absent when $n_4=0$.
Since the mass term for $\tr(\hat A_i^2)$ and $a_i$
do not have illuminating expressions in general  if we eliminate
$\tr (A_{44})$,  we examine these terms for each case.
As it turns out, however,  we see that
the above choice~(\ref{N8_CP_dyonic_canscalars_tr})
always leads to the diagonal mass matrix for the trace part.

Next, we shall determine the eigenvalue of $U$.
To this end, we classify the basis of self-dual 4-form $U$
in terms of the degeneracy of eigenvalues
$(\lambda_i, \kappa_i)$.  Let us define
\begin{align}
 \vec \ell := (\ell_1,..., \ell_m), \qquad \ell_i \le n_i \,, \qquad
\sum_i^m \ell_i=4 \qquad m\le 4\,,
\end{align}
where $\ell_i$'s are nonnegative integers.
The multiplicities belonging to the same $\vec \ell$
are given by
\begin{subequations}
\begin{align}
{}_{n_1}C_{\ell_1}\times \cdots \times {}_{n_m}C_{ \ell_m}: &\quad
\ell_1 >n_1/2 \quad {\rm or} \quad \ell_1=n_1/2, ~\ell_2 >n_2/2 \quad
 {\rm or} ,... \,,\nonumber \\
& \quad \ell_1 =n_1/2 , ~\ell_2=n_2/2 , ...,~ \ell_m >n_m/2 \,.\\
\frac{1}{2}
{}_{n_1}C_{n_1/2}\times \cdots \times {}_{n_m}C_{n_m/2}: &\quad
\ell_i= n_i/2 ~~(i=1,...,m)\,.
\end{align}
\end{subequations}
We can easily verify
\begin{align}
 (1+\star )(\theta^r{} _{[a}\theta ^s{} _{b}+\xi^r{} _{[a}\xi^s{}
 _{b})U_{cd]rs}&=2 \mu _{\vec \ell }
 U_{abcd} \,,
\end{align}
where
\begin{align}
\mu_{\vec \ell} = \frac{1}{24}& \left[\left(
\sum_i \ell_i\lambda_i\right)^2 +\left(\sum_i (n_i-\ell_i)\lambda_i\right)^2
+\left(\sum_i \ell_i \kappa_i \right)^2
+\left(\sum_i(n_i-\ell_i)\kappa_i\right)^2-\sum_i n_i (\lambda_i^2
 +\kappa_i^2)  \right]\,.
\end{align}
Hence  we obtain  the mass eigenvalues
\begin{align}
M_{(2)}^2 =\frac{1}{6} m_{[\vec \ell]}^2 U\cdot U \,,
\end{align}
with
\begin{align}
m_{[\vec \ell]}^2= \frac{g^2}{32}& \left[
2\sum_i n_i (\lambda_i^2+\kappa_i^2)-\left(\sum_i \ell_i
 \lambda_i\right)^2
-\left(\sum_i (n_i-\ell_i)\lambda_i \right)^2
\right. \nonumber \\& \left. ~~
-\left(\sum_i \ell_i \kappa _i\right)^2
-\left(\sum_i (n_i-\ell_i)\kappa _i \right)^2
\right]\,,
\label{N8_CP_dyonic_mass_U}
\end{align}
which is specified by nonnegative integers $\ell_i$ satisfying
\begin{align}
0\le \ell_i \le n_i \,, \qquad
\sum_i^m \ell_i =4 \,.
\end{align}
Since the kinetic term for scalars is given
by~(\ref{N8_CP_electric_Skin}),
$m_{[\vec \ell]}^2$ denotes the canonical mass eigenvalues.

We are now armed with necessary tools to demonstrate
mass spectra in the dyonic case.

\bigskip\noindent{(I) $\theta \propto \xi ^{-1}$. }
Let us begin with the $\theta \propto \xi ^{-1}$ case.

\bigskip\noindent{(i) $n_i=2$.}
This case corresponds to the Minkowski vacua of ${\rm SO}(6,2)$ gauging,
which spontaneously breaks down to
${\rm SO}(2)\times{\rm SO}(2)\times{\rm SO}(2)\times{\rm SO}(2)$.
Taking the $\theta $ tensor as~(\ref{N8_CP_SO62_theta}),
equations~(\ref{N8_CP_dyonic_M1sq_dec}) and (\ref{N8_CP_dyonic_canscalars_tr})
yield
\begin{align}
&m^2_{(\mathbf 1,\mathbf 1,\mathbf 1)(\times 3)} =0 \,, \qquad  m_{*(\times 24)}^2=0\quad
 :(\mathbf 2,\mathbf 2,\mathbf 1,\mathbf 1)+\cdots \,, \nonumber \\
& m^2_{i(\times 8)} =\frac{g^2}{16 r^2s^2t^2} \times
\left\{
\begin{array}{ll}
4 st(r-s)(r-t)(1+r^2st)\,, & :(\mathbf 2,\mathbf 1,\mathbf 1,\mathbf 1)\\
4 rt(s-r)(s-t)(1+rs^2 t)\,, & :(\mathbf 1,\mathbf 2,\mathbf 1,\mathbf 1)\\
4 rs(r-t)(s-t)(1+rst^2)\,, & :(\mathbf 1,\mathbf 1,\mathbf 2,\mathbf 1)\\
4(1+r^2 st)(1+r s^2t )(1+rst^2)\,, ~&:(\mathbf 1,\mathbf 1,\mathbf 1,\mathbf 2)
\end{array}
\right. \,.
\label{N8_CP_dyonic_SO62_mass_S}
\end{align}
From~(\ref{N8_CP_dyonic_mass_U}),
the mass eigenvalues for pseudoscalars are given by
\begin{align}
& m^2_{(\mathbf 1,\mathbf 1,\mathbf 1, \mathbf 1)(\times 3)}=0 \qquad :
~
[2,2,0,0]+[2,0,2,0]+[2,0,0,2]\,,
 \nonumber \\
& m^2_{[\vec \ell]} =\frac{g^2}{16 r^2s^2 t^2}
\times \left\{
\begin{array}{ll}
t^2(r-s)^2 (1+r^2s^2) \,,& :(\mathbf 2,\mathbf 2,\mathbf 1,\mathbf 1)=[1,1,2,0]_{(\times 4)}  \\
s^2 (r-t)^2 (1+r^2 t^2)  \,,&:(\mathbf 2,\mathbf 1,\mathbf 2,\mathbf 1)=[1,2,1,0]_{(\times 4)} \\
r^2 (s-t)^2 (1+s^2t^2) \,,& :(\mathbf 1,\mathbf 2,\mathbf 2,\mathbf 1)=[2,1,1,0]_{(\times 4)} \\
(1+s^2 t^2)(1+s r^2 t)^2 \,,&:(\mathbf 2,\mathbf 1,\mathbf 1,\mathbf 2)=[1,2,0,1]_{(\times 4)} \\
(1+s^2 r^2)(1+s r t^2)^2 \,,& :(\mathbf 1,\mathbf 1,\mathbf 2,\mathbf 2)=[2,0,1,1]_{(\times 4)} \\
(1+r^2 t^2)(1+s^2 r t)^2  \,,& :(\mathbf 1,\mathbf 2,\mathbf 1,\mathbf 2)=[2,1,0,1]_{(\times 4)} \\
(1+s^2 t^2 )(1+r^2 s^2)(1+t^2 r^2)\,,~  & :(\mathbf 2,\mathbf 2,\mathbf 2,\mathbf 2)^+=[1,1,1,1] _{(\times 8)}
\end{array}
\right.\,.\label{N8_CP_dyonic_SO62_mass_U}
\end{align}
It is emphasized that mass eigenvalues for the pseudoscalars are
always nonnegative, whereas those for scalars are not.
This implies that vacua of generic ${\rm SO}(6,2)$
gauging are unstable~\cite{Dall'Agata:2012cp}.
In the special case where $(r,s,t)$ are pairwise equal,
all mass eigenvalues become nonnegative,
corresponding to the stable Minkowski vacua
found in~\cite{DI}.

\bigskip\noindent{(ii) $p_3=-p_4(\ne 0)$.}
This case corresponds to the ${\rm SO}(4,4)$ dS vacua.
From (\ref{N8_CP_ET_m1_electric}) and
(\ref{N8_CP_ET_m2_electric}), we obtain
\begin{align}
&m_{0(\mathbf 1,\mathbf 1)}^2 =-2 V_c \,, \qquad
m_{1(\mathbf 9,\mathbf 1)}^2 =m_{2(\mathbf 1, \mathbf 9)}^2
=2 V_c \,, \qquad m_{*(\mathbf 4,\mathbf 4)}^2 =0 \,,\nonumber \\
&
m^2 _{[\ell=2](\times 18)}=
m^2_{(\mathbf  6, \mathbf 6)^+}=2 V_c \,, \quad
m^2 _{[\ell=3](\times 16)}=
m^2_{(\mathbf 4,\mathbf 4)}= V_c \,, \quad
m^2 _{[\ell=4](\times 1)}=m^2_{(\mathbf 1,\mathbf 1)}=-2 V_c \,.
\end{align}
The tachyonic modes emerge from the ${\rm SO}(4)\times {\rm SO(4)}$
invariant sector.
These spectra agree with the electric case.

\bigskip\noindent{(iii) $p_3=-p_4(\ne 0)$.}
In this case the cosmological constant is nonvanishing.
Using the expression~(\ref{n8_CP_dyonic_III_pot}),
the mass term~(\ref{N8_CP_dyonic_M1sq_dec}) for $S$  considerably simplifies to
\begin{align}
\sum_{i,j}\mu_{ij}^2\tr (A_{ii}) \tr (A_{jj})
=-2V_c  \sum_{i=1}^3 a_i^2  \,, \qquad
m_{i}^2 = \frac{4}{n_i-2} V_c \,.
\label{N8_CP_dyonic_III_Smass}
\end{align}
where canonical scalars $a_i$ were defined in (\ref{N8_CP_dyonic_canscalars_tr}).
As we already explained,  $A_{ij}~(i<j)$ do not contribute to the mass term.
In the case of dS vacua ($V_c>0$), the scalars $a_i$ are always tachyonic.
The above equation  exhibits that the mass spectrum is
determined only by $n_i$, which controls the residual gauge symmetry.
It also illustrates that the mass spectrum is completely
independent of the deformation parameter $s$ when normalized by the
cosmological constant.
In other words the ``slow-roll matrix''
$\eta =\partial_\rho \partial_\sigma V/V$
does not depend on $s$, albeit the change of the value of
potential. We can also find that the mass spectrum for $U$
shares this property.

\bigskip\noindent{(8,0,0,0):}
This case corresponds to the  maximally supersymmetric
AdS vacua of the ${\rm SO}(8)$ gauging.
The $\theta $ tensor and the potential are given by (\ref{N8_CP_dyonic_theta_8080}).
The formulas (\ref{N8_CP_ET_m1_electric}) and
(\ref{N8_CP_ET_m2_electric}) yield
\begin{align}
m^2_{1(\mathbf{35})} =-\frac 23 |V_c| \,,\qquad
m^2_{[ \ell =4](\mathbf{35})} =-\frac 23 |V_c| \,.
\end{align}
Since all of these mass eigenvalues are above
the Breitenlohner-Freedman bound~\cite{Breitenlohner:1982bm},
the vacuum is stable as expected from supersymmetry.

\bigskip\noindent{(7,1,0,0):}
This gives AdS vacua of ${\rm SO}(8)$ and ${\rm SO}(7,1)$
gaugings with an ${\rm SO}(7)$ residual gauge symmetry.
The $\theta $ tensor and the potential are given by
(\ref{N8_CP_dyonic_theta_7100}).
From  (\ref{N8_CP_ET_m1_electric}) and
(\ref{N8_CP_ET_m2_electric}),
the mass spectrum now reads
\begin{align}
m_{0(\mathbf 1)}^2  & = 2 |V_c| \,, \qquad
m_{1(\mathbf{27})}^2   = -\frac 45 |V_c| \,, \qquad
m_{[\ell =4] (\mathbf{35})}^2 =-\frac 25|V_c| \,.
\label{N8_CP_dyonic_7110U}
\end{align}
$m_{*(\mathbf 7)}^2=0$ correspond to the
Nambu-Goldstone fields.

\bigskip\noindent{(5,3,0,0):}
This case corresponds to the  ${\rm SO}(5,3)$ gauging dS vacua
with a residual
gauge symmetry ${\rm SO}(5)\times {\rm SO}(3)$.
The $\theta $ tensor and the potential are given by (\ref{N8_CP_dyonic_theta_5300}).
Equations (\ref{N8_CP_ET_m1_electric}) and
(\ref{N8_CP_ET_m2_electric}) yield
\begin{align}
m_{0(\mathbf 1, \mathbf 1)}^2  & = -2 V_c \,, \qquad
m_{1(\mathbf{14},\mathbf 1)}^2   = \frac 43 V_c \,, \qquad
m_{2(\mathbf 1, \mathbf 5)}^2   =  4 V_c \,, \nonumber \\
m_{[\ell =3] (\times 30)}^2 &=m^2 _{(\mathbf{10}, \mathbf 3) }=2 V_c \,, \qquad
m_{[\ell =4] (\times 5)}^2 =m^2_{(\mathbf 5, \mathbf 1)}=-\frac 23 V_c \,.
\end{align}
$m_{*(\mathbf 5,\mathbf 3)}^2=0$ correspond to the
Nambu-Goldstone fields.

Let us compare this with the result in~\cite{Hull:1984ea,Ahn:2002qga},
where the ${\rm SO}(5)$ invariant scalars were
analyzed.
The corresponding potential was shown to be
\begin{align}
V=\frac{g^2}{8}\left(\frac{u^3v^3}{w^3}+\frac{10uv}{w}-2uvw^3
+\textrm{two cyclic perm.}-\frac{15}{uvw}\right)\,,
\end{align}
where $u$, $v$ and $w$ are ${\rm SO}(5)$ invariant (unnormalized)
scalars. The dS vacuum $V_c=2\times 3^{1/4}g^2$ exists at
$u=v=w=3^{-1/4}$ (observe that the normalization of $g$ is different
from the present paper).   From the expression of scalar kinetic term
in reference~\cite{Hull:1984ea,Ahn:2002qga},
one can find the canonically normalized scalars $\Phi_i~(i=1,2,3)$
as
\begin{align}
 \Phi_1=\sqrt{\frac{5}{6}} \ln (uvw) \,, \qquad
\Phi_2 =\frac{1}{\sqrt 6} \ln \left(\frac{v^2}{uw}\right) \,, \qquad
\Phi_3 =\frac{1}{\sqrt 2} \ln \left(\frac{w}{u}\right)\,.
\end{align}
Then the mass eigenvalues are given by
\begin{align}
m_{\Phi_1}^2 =-2 V_c\,, \qquad
m_{\Phi_2}^2=m_{\Phi_3}^2=4 V_c \,.
\end{align}
These spectra coincide with the result obtained above,
where ${\rm SO}(5)$ invariant scalars descend from
${\rm Tr}(A_{11})$ and  ${\rm Tr}(\hat A_{2}^2)$.

\bigskip\noindent{(6,0,1,1):}
This case gives the AdS vacua of ${\rm SO}(8)$
and  ${\rm SO}(7,1)$ gaugings with
the residual gauge symmetry ${\rm SO}(6)$.
The $\theta $ tensor and the potential are
given by (\ref{N8_CP_dyonic_theta_6100}).
Now the formula (\ref{N8_CP_dyonic_III_Smass}) can be applied
to give
\begin{align}
m_{0(\mathbf 1)\times 2}^2 =2 |V_c| \,, \qquad
m^2_{1(\mathbf{20})} =- |V_c| \,, \qquad
m^2_{*({\bf 6})(\times 2)}=m^2_{*(\mathbf 1)}=0 \,.
\end{align}
For the $U$ field, equation~(\ref{N8_CP_dyonic_mass_U}) yields
\begin{align}
m^2_{[4,0,0](\mathbf{15})}= 0 \,, \qquad
m^2 _{[3,1,0] (\mathbf{ 20})}=-\frac 14 |V_c| \,.
\label{N8_CP_dyonic_6110U}
\end{align}

\bigskip\noindent{(5,1,1,1):}
We have the AdS critical point
of ${\rm SO}(7,1)$ gauging with the residual gauge symmetry ${\rm SO}(5)$.
The $\theta $ tensor and the potential are
given by (\ref{N8_CP_dyonic_theta_6100}).
Using (\ref{N8_CP_dyonic_III_Smass}) we obtain
\begin{align}
m^2_{0({\bf 1})(\times 3)} =2|V_c| \,, \qquad
m^2_{1 ({\bf 14})} =-\frac 43 |V_c| \,, \qquad
m_{*({\bf 5})(\times 3)}^2=m^2 _{*({\bf 1})(\times 3)}=0\,.
\end{align}
From (\ref{N8_CP_dyonic_mass_U}), the mass eigenvalues for $U$  are given by
\begin{align}
m^2_{[3,1,0,0](\mathbf{10})}=m^2_{[3,0,1,0](\mathbf{
 10})}=m^2_{[3,0,0,1](\mathbf{ 10})}=0 \,, \qquad
m_{[4,0,0,0](\mathbf 5)}^2 =\frac 23 |V_c| \,.
\label{N8_CP_dyonic_5111U}
\end{align}

\bigskip\noindent{(II) $\theta \xi =0$.}
We next turn to study the $\theta \xi =0$ case.

\bigskip\noindent{(i-$a$)}
For the
${\rm SO}(4)\times {\rm SO}(2,2)\ltimes \mathbb T^{16}$
gauging with a residual symmetry
${\rm SO}(4)\times {\rm SO}(2)\times {\rm SO}(2)$,
$\theta $ and $\xi $ are given by (\ref{N8_CP_mag_deg_thetaxi_case1})
and the Minkowski vacua are realized.
The mass spectra are given by
\begin{align}
&m^2_{0(\mathbf 1,\mathbf 1,\mathbf 1)(\times 3)}=0\,, \qquad
m_{*(\times 24)}^2 =0\quad :(\mathbf 2,\mathbf 2, \mathbf 1,\mathbf
 1)+\cdots  \,, \nonumber \\
&m^2_{i}=\frac{g^2}{16s^2t^2}\times \left\{
\begin{array}{ll}
-4s^2 t^2(1-s^2)\,, & :(\mathbf 2,\mathbf 1,\mathbf 1,\mathbf 1 )\\
4 t^2(1-s^2)\,,~ & :(\mathbf 1,\mathbf 2,\mathbf 1,\mathbf 1 )\\
4s^2t^2(1+t^2)\,, & :(\mathbf 1,\mathbf 1,\mathbf 2,\mathbf 1 )\\
4s^2(1+t^2)\,, & :(\mathbf 1,\mathbf 1,\mathbf 1,\mathbf 2 )
\end{array}
\right.  \,.  \label{N8_CP_dyonic_SO4xSO22_mass_S}
\end{align}
For the pseudoscalars, we have
\begin{align}
& m^2_{(\mathbf 1, \mathbf 1,\mathbf 1, \mathbf 1)(\times 3)}=0 \qquad :~[2,2,0,0]+[2,0,2,0]+[2,0,0,2]
 \,, \nonumber \\
&m^2_{[\vec \ell]} = \frac{g^2}{16s^2 t^2} \times\left\{
\begin{array}{ll}
t^2 (1+s^2t^2)\,, & :(\mathbf 1,\mathbf 2,\mathbf 2,\mathbf 1)=[2,1,1,0]_{(\times 4)}\\
(s^2+t^2)\,, & :(\mathbf 1,\mathbf 2,\mathbf 1,\mathbf 2)=[2,1,0,1]_{(\times 4)} \\
s^2 (1+s^2 t^2)\,, & :(\mathbf 2,\mathbf 1, \mathbf 1, \mathbf 2)=[1,2,0,1]_{(\times 4)} \\
s^2t^2 (s^2+t^2)\,, & :(\mathbf 2, \mathbf 1,\mathbf 2 ,\mathbf 1)=[1,2,1,0]_{(\times 4)} \\
t^2 (1-s^2)^2\,, & :(\mathbf 2, \mathbf 2, \mathbf 1, \mathbf 1)=[1,1,2,0]_{(\times 4)} \\
s^2 (1+t^2)^2 \,, & :(\mathbf 1, \mathbf 1, \mathbf2, \mathbf 2)=[2,0,1,1]_{(\times 4)} \\
(s^2+t^2)(1+s^2t^2)\,,~ & :(\mathbf 2, \mathbf 2,\mathbf 2,\mathbf 2)^+=[1,1,1,1]_{(\times 8)}
\end{array}
\right.\,.
\label{N8_CP_dyonic_SO4xSO22_mass_U}
\end{align}
The pseudoscalars are all stable, whereas the scalars
are unstable unless $s=1$.

\bigskip\noindent{(i-$b$) }
For the   ${\rm SO}(2)\times {\rm SO}(2)\ltimes \mathbb T^{20}$
gauging with a residual symmetry
 ${\rm SO}(2)\times {\rm SO}(2)$,
$\theta $ and $\xi $ are given by (\ref{N8_CP_mag_deg_thetaxi_case2})
and the Minkowski vacua are realized. The mass spectra are
\begin{align}
&m^2_{0(\mathbf 1,\mathbf 1)(\times 2)}=0 \,, \qquad
m_i^2 =\frac{1}{4}g^2 \times \left\{
\begin{array}{ll}
1 &:(\mathbf 2,\mathbf 1) \\
s^2 & :(\mathbf 1, \mathbf 2)\\
0 & :(\mathbf 1, \mathbf 1)_{(\times 9)}
\end{array}
\right.\,,\nonumber \\&
 m_{*(\times 20)}^2=0 \quad :(\mathbf 2, \mathbf 1)_{(\times 4)}+
(\mathbf 1, \mathbf 2)_{(\times 4)}+(\mathbf 2, \mathbf 2)\,,
\label{N8_CP_dyonic_SO2xSO2_mass_S}
\end{align}
and
\begin{align}
&m^2_{(\mathbf 1,\mathbf 1)(\times 7)}=0 \quad :[2,2,0]+[2,0,2]_{(\times
 6)}\,, \nonumber \\
&m^2_{[\vec \ell]}
=\frac{1}{16}g^2 \times \left\{
\begin{array}{ll}
1\,, &(\mathbf 2, \mathbf 1)_{(\times 4)}=[1,2,1]_{(\times 8)} \\
s^2 \,,& (\mathbf 1,\mathbf 2)_{(\times 4) }=[2,1,1]_{(\times 8 )} \\
(1+s^2)\,,~ & (\mathbf 2,\mathbf 2)^+_{(\times 6)} = [1,1,2] _{(\times 12)}
\end{array}
\right.
\label{N8_CP_dyonic_SO2xSO2_mass_U}
\end{align}
Hence these vacua are stable.

\bigskip\noindent{(ii) $n_i \ne 2$. }
The AdS vacua are realized in this family.
Using (\ref{N8_CP_dyonic_canscalars_tr}),
(\ref{N8_CP_dyonic_deg_CPcond}) and (\ref{N8_CP_pot_deg}) with
$\lambda_3=\lambda_4=\kappa_1=\kappa_2=0$,
the mass spectrum of $S$ is given by
\begin{align}
M_{(1)}^2 =|V_c|  \left[
-\sum_{i=1}^4 \frac{4}{n_i-2} \tr (\hat A_i^2)
+ 2 \sum_{i=1}^3  a_i^2
\right] \,.
\end{align}
It is gratifying that this expression accords precisely with
(\ref{N8_CP_dyonic_III_Smass}), for which
$\xi \propto \theta^{-1}$. $A_{ij}$ are always massless
irrespective of the gaugings.
The above equation also confirms that
the mass spectrum for $S$ is only dependent on $n_i$'s, i.e.,
the residual gauge symmetry only. The same is
true for the $U$ field.

\bigskip\noindent{(7,0,1,0):}
This corresponds to the ${\rm SO}(7) \ltimes \mathbb T^7$ gauging with an
${\rm SO}(7)$ remaining symmetry, where
$\theta $ and $\xi $ are given by~(\ref{N8_CP_dyonic_deg_theta_7010}).
We obtain
\begin{align}
 m_{0 (\mathbf 1)}^2=2 |V_c| \,, \qquad
 m_{1({\bf 27})}^2 =-\frac{4}{5}|V_c| \,, \qquad
 m^2_{*(\mathbf 7)}=0 \,,
\end{align}
and
\begin{align}
 m^2_{[4,0](\mathbf{35})} =-\frac 25 |V_c| \,.
\end{align}

\bigskip\noindent{(6,1,0,1):}
We have the  ${\rm SO}(7) \ltimes \mathbb T^7$ gauging with an
${\rm SO}(6)$ remaining symmetry, where
$\theta $ and $\xi $ are given by
(\ref{N8_CP_dyonic_deg_theta_6110}). We obtain
\begin{align}
 m_{0 (\mathbf 1)(\times 2)}^2=2 |V_c| \,, \qquad
 m_{1(\mathbf{20})}^2 =-|V_c| \,, \qquad
 m^2_{*(\mathbf 6)(\times 2)}=m^2 _{*({\bf 1})}=0 \,,
\end{align}
and
\begin{align}
 m^2_{[3,1,0](\mathbf{20})}=-\frac 14 |V_c| \,, \qquad
 m^2_{[4,0,0](\mathbf{ 15})}=0 \,.
\end{align}

\bigskip\noindent{(6,0,1,1):}
We have the ${\rm SO}(6)\times {\rm SO}(1,1)\ltimes \mathbb T^{12}$ gauging
with an ${\rm SO}(6)$ remaining symmetry,
where $\theta $ and $\xi $ are given by
(\ref{N8_CP_dyonic_deg_theta_6011}).  We obtain
\begin{align}
 m_{0 (\mathbf 1)(\times 2)}^2=2 |V_c| \,, \qquad
 m_{1(\mathbf{20})}^2 =-|V_c| \,, \qquad
 m^2_{*(\mathbf 6)(\times 2)}=m^2 _{*({\bf 1})}=0 \,,
\end{align}
for scalars and
\begin{align}
  m^2_{[3,1,0](\mathbf{20})}=-\frac 14 |V_c| \,, \qquad
 m^2_{[4,0,0](\mathbf{ 15})}=0 \,.
\end{align}
for pseudoscalars.
These are the same as (6,1,1,0) type since the residual
gauge symmetries are equivalent.

\bigskip\noindent{(5,1,1,1):}
We have the ${\rm SO}(6)\times {\rm SO}(1,1)\ltimes \mathbb T^{12}$ gauging
with an ${\rm SO}(6)$ remaining symmetry,
where $\theta $ and $\xi $ are given by
(\ref{N8_CP_dyonic_deg_theta_5111}). We obtain
\begin{align}
 m_{0(\mathbf 1) (\times 3)}^2=2 |V_c| \,, \qquad
 m_{1(\mathbf{ 14})}^2 =-\frac{4}{3}|V_c| \,, \qquad
 m^2_{*({\bf 5})(\times 3)}=m^2_{*(\mathbf 1)(\times 3)}=0 \,,
\end{align}
and
\begin{align}
m^2_{[3,1,0,0](\mathbf{10})} =m^2_{[3,0,1,0](\mathbf{10})}
=m^2_{[3,0,0,1](\mathbf{10})}=0 \,, \qquad
m^2_{[4,0,0,0](\mathbf{5})} =\frac 23 |V_c| \,.
\end{align}

We enumerate the result obtained in this section in table~\ref{table:N8_CP_dyonic}.
Except for the maximally supersymmetric AdS vacua,
supersymmetries are broken completely. One can inspect that
at the nonsupersymmetric AdS vacua, the  $S$ field
does not respect the Breitenlohner-Freedman bound, implying the linear
instability of these vacua.
For the Minkowski vacua, the mass spectrum for $S$ is not necessarily
positive, implying the instability.

\begin{table}
\label{table:N8_CP_dyonic}
\begin{center}
{\small
\begin{tabular}[t]{l||cccc}
Gauging   & $G_{\rm reg}$ &$\Lambda $ &$m^2_{S}$ & $m^2_U$
\\ \hline \hline
${\rm SO}(4,4)$  &${\rm SO}(4)\times {\rm SO(4)}$ & dS &
$-2_{(\times 1)}$, $2_{(\times 18)}$, $0_{(\times 16)}$ &
$2_{(\times 18)}, 1_{(\times 16)}, -2_{(\times 1)}$\\ \hline
${\rm SO}(5,3)$ &${\rm SO}(5)\times {\rm SO(3)}$ & dS
& $-2_{(\times 1)}$, $\frac 43_{(\times 14)}$, $4_{(\times 5)}$, $0_{(\times 15)}$
 & $2_{(\times 30)}$, $-\tfrac 23_{(\times 5)}$ \\ \hline\hline
${\rm SO}(8)$  &  &
 &  &
 \\
${\rm SO}(7,1)$  &${\rm SO}(7)$ &
AdS & $2_{(\times 1)}$, $-\tfrac 45_{(\times 27)}$, $0_{(\times 7)}$ &
$-\tfrac 25_{(\times 35)}$
\\
${\rm SO}(7)\ltimes \mathbb T^7$  & &
 & &
 \\ \hline
${\rm SO}(8)$ & \multirow{4}*{${\rm SO(6)}$} &
\multirow{4}*{AdS }&\multirow{4}*{$-1_{(\times 20)}$,
$2_{(\times 2)}$, $0_{(\times 13)}$
}&\multirow{4}*{$0_{(\times 15)} $, $-\tfrac 14 _{(\times 20)}$} \\
${\rm SO}(7,1)$ &  & &  &
\\
${\rm SO}(7)\ltimes \mathbb T^7$ &&&& \\
${\rm SO}(6)\times{\rm SO}(1,1)\ltimes \mathbb T^{12}$ &&&& \\ \hline
${\rm SO}(7,1)$ & \multirow{2}*{${\rm SO(5)}$} &
\multirow{2}*{AdS }&\multirow{2}*{$-\tfrac 43_{(\times 14)}$,
	     $2_{(\times 3)}$, $0_{(\times 18)}$}
&\multirow{2}*{$\tfrac 23 _{(\times 5)}$, $0_{(\times 30)}$} \\
${\rm SO}(6)\times{\rm SO}(1,1)\ltimes\mathbb T^{12}$ &  & &  &
\\ \hline\hline
${\rm SO}(6,2)$ & ${\rm SO}(2)^4$ & Mink. &
	     Eq.~(\ref{N8_CP_dyonic_SO62_mass_S})
& Eq.~(\ref{N8_CP_dyonic_SO62_mass_U}) \\\hline
${\rm SO}(4)\times{\rm SO}(2,2)\ltimes \mathbb T^{16}$
& ${\rm SO}(2)^4$ & Mink. & Eq.~(\ref{N8_CP_dyonic_SO4xSO22_mass_S})
& Eq.~(\ref{N8_CP_dyonic_SO4xSO22_mass_U}) \\\hline
${\rm SO}(2)\times{\rm SO}(2)\ltimes \mathbb T^{20}$
& ${\rm SO}(2)^2$ & Mink. & Eq.~(\ref{N8_CP_dyonic_SO2xSO2_mass_S})
& Eq.~(\ref{N8_CP_dyonic_SO2xSO2_mass_U}) \\
\hline\hline
\end{tabular}
}
\caption{Mass spectrum for dyonic gaugings.
Except for the Minkowski vacua, mass eigenvalues are
normalized by the absolute value of cosmological constant.
Supersymmetries are completely broken.}
\end{center}
\end{table}

\section{Concluding remarks}
\label{sec:conclusion}

We have studied the critical points and their
mass spectra in maximal gauged supergravity.
Although the maximal supergravity is not entitled
to a unified framework for gauge interactions,
scanning vacua in this theory certainly serves as a foundation for
the realistic construction of string vacua, due to the restrictive
property of maximal supergravity.
In particular, the result of this vacuum search can have significant
implications to the construction of inflationary universe models
on the base of string/M theory, because the maximal gauged supergravity
may describe the gravity sector very well, including non-perturbative
effects in the 10/11-dimensional framework.
In addition it is also useful for the phenomenological applications
to the condensed matter physics.

Utilizing the fact that the scalar fields parametrize the
homogeneous space, we can analyze the
70 scalar mass spectrum at the origin of scalar space
as argued in~\cite{DI,Dibitetto:2011gm}.
Specializing to the cases in which the gauge group is embedded into the standard ${\rm SL}(8, \mathbb R)$ subgroup of $E_{(7(7)}$,
we were able to enumerate all the possible vacua in this class corresponding to critical points
that can be mapped to the origin by a transformation in the standard ${\rm SL}(8, \mathbb R)$ group.
We also developed a new formulation which
allows us to obtain the analytic expression of mass spectra
in terms of eigenvalues of the embedding tensor.
We established an interesting structure
about the moduli space of vacua:  when the
cosmological constant is nonvanishing,
the mass spectrum is only sensitive
to the residual gauge symmetry at the vacua.
Namely, the mass spectra for the \SL-type gaugings have to coincide among the
different theories as long as their residual gauge symmetries
are identical. This resolved the issue which remained open in~\cite{DI}.

In some cases of dyonic gaugings,
we are left with a deformation parameter $s$.
It turns out that the mass spectrum is nevertheless
insensitive to the parameter $s$ in units of the cosmological constant.
This means that ${\rm SO}(4,4)$ and ${\rm SO}(5,3)$ dS maxima do not
provide sufficient $e$-foldings in the standard potential-driven inflation
scenario  even in the deformed theory, since the
slow-roll parameter $\eta $ is of order unity.
We can also verify that the fraction of residual supersymmetries
is not dependent on the deformation parameter, i.e.,
all vacua except the maximally supersymmetric AdS totally
break supersymmetries.

We have also shown that the generic Minkowski vacua found in this paper
do not have stable mass spectra unless the
remaining continuous parameters are finely tuned.
This is consistent with the result in ~\cite{Dall'Agata:2012cp}.

The obvious next step is to explore the vacuum classifications
for gaugings contained in other subgroup of $E_{7(7)}$, such as
$E_{6(6)}$  and ${\rm SU}^*(8)$.
We believe that the techniques developed in this paper
could be used in other frames.
It is interesting to see whether the
characteristic features exposed here are universal, i.e.,
whether the mass spectrum is insensitive to the deformation parameter and
only dependent on the residual gauge symmetry.

Another possible future work is to
work out inflationary models in the maximal theory.
As we have demonstrated systematically,
gaugings contained in the \SL~frame fail to have
stable dS vacua and the slow-roll condition is never satisfied.
Even though a simple hill-top type inflation does not work,
there remains a possibility for a realization of sufficient inflation
around these dS saddle points with the aid of other fields.
We have 35 scalars and 35 pseudoscalars, which may be able to realize
quasi-dS phase if flux is turned on appropriately.
We will report this issue elsewhere.

\section*{Acknowledgements}

We thank all participants of the workshop ExDiP2012 ``Superstring
Cosmophysics,'' in particular Gianguido Dall'Agata for valuable discussions.
This work is supported in part by the MEXT
Grant-in-Aid for Scientific Research on Innovative Areas
No. 21111006 and the JSPS Grant-in-Aid for Scientific Research (A) (22244030).

\end{document}